\documentclass[
 reprint,
 amsmath,amssymb,
 aps, pre,
]{revtex4-2}

\usepackage{appendix}
\usepackage{graphicx}
\usepackage{dcolumn}
\usepackage{bm}
\usepackage[hidelinks]{hyperref}
\usepackage[capitalise]{cleveref}
\usepackage{lipsum}
\usepackage[dvipsnames]{xcolor}
\usepackage{amsfonts}
\usepackage{yfonts}
\usepackage{nicematrix}
\NiceMatrixOptions{cell-space-limits = 2pt}

\usepackage[english]{babel}
\usepackage{amsthm}
\theoremstyle{definition}
\newtheorem{lemma}{Lemma}

\DeclareFontFamily{U}{futm}{}
\DeclareFontShape{U}{futm}{m}{n}{
  <-> s * [.97534] fourier-bb 
  }{}
\DeclareMathAlphabet{\mathbbs}{U}{futm}{m}{n}

\begin{document}

\preprint{APS/123-QED}

\title{
Unified Linear Fluctuation-Response Theory Arbitrarily Far from Equilibrium
}

\author{Jiming Zheng}
\email{jiming@unc.edu}
\affiliation{Department of Chemistry, University of North Carolina-Chapel Hill, NC}
\author{Zhiyue Lu}
\email{zhiyuelu@unc.edu}
\affiliation{Department of Chemistry, University of North Carolina-Chapel Hill, NC}

\date{\today}

\begin{abstract}
Understanding how systems respond to external perturbations is a fundamental challenge in physics, particularly for non-equilibrium and non-stationary processes. The fluctuation-dissipation theorem provides a complete framework for near-equilibrium systems, and various bounds have recently been reported for specific non-equilibrium regimes. Here, we present an exact response equality for arbitrary Markov processes that decompose system response into spatial correlations of local dynamical events. This decomposition reveals that response properties are encoded in correlations between transitions and dwelling times across the network, providing a natural generalization of the fluctuation-dissipation theorem and recently developed non-equilibrium linear response relations. Our theory unifies existing response bounds, extends them to time-dependent processes, and reveals fundamental monotonicity properties of the tightness of multi-parameter response inequalities. Beyond its theoretical significance, this framework enables efficient numerical evaluation of response properties from sampling unperturbed trajectories, offering significant advantages over traditional finite-difference approaches for estimating response properties of complex networks and biological systems far from equilibrium.

\end{abstract}

\maketitle

\section{Introduction}

The measurement and prediction of system response to external perturbations stands as a fundamental challenge in physics. The celebrated fluctuation-dissipation theorem (FDT) \cite{kubo1957statistical} revealed that near equilibrium, a system's response to weak perturbations is fully determined by its equilibrium fluctuations. This principle is captured by the relation $\frac{\partial \langle Q \rangle}{\partial \lambda} = \beta \operatorname{Cov}_{\text{eq}} (Q, \Lambda)$, showing that an observable $Q$'s response to a parameter perturbation $\lambda$ is encoded in its equilibrium correlation with the parameter's conjugate variable $\Lambda$. This remarkable result demonstrates that a system's response properties can be accessed through equilibrium measurements alone, without the need for perturbative experiments.

Non-equilibrium systems, ubiquitous in nature and technology, present a broader challenge for response theory. Pioneered by Agarwal \cite{agarwal1972fluctuation}, the linear response theory is developed near non-equilibrium steady states (NESS) through the lens of the Green's function method \cite{wu2020generalized,zhang2021quantum,feng2011potential,marconi2008fluctuation}. Building upon the formulation of fluctuation theorems and trajectory ensemble methods \cite{jarzynski1997nonequilibrium,evans1993probability,dellago1998transition,dellago1998efficient,bolhuis2002transition,touchette2009large,chetrite2013nonequilibrium,seifert2012stochastic,bertini2015macroscopic,maes2020frenesy}, researchers developed a trajectory-level formulation of stochastic response theory. Especially, Seifert and Speck \cite{seifert2010fluctuation} generalized the Agarwal form of linear response theory to NESS under the view of stochastic thermodynamics, systematically addressing the role of trajectory entropy in the linear response region around NESS. Baiesi, Maes, and Wynants \cite{baiesi2009nonequilibrium,baiesi2013update,maes2020response} further provided a general linear response theory using trajectory ensembles, explicitly decomposing response functions into entropy-production and dynamical-activity (``frenetic'') contributions. More recently, significant progress has been made on the fluctuation-response inequalities (FRIs) \cite{dechant2020fluctuation,zheng2024universal,kwon2024fluctuation,ptaszynski2024dissipation,aslyamov2025nonequilibrium,ptaszynski2024nonequilibrium,ptaszynski2025nonequilibrium,liu2025dynamical}. In particular, Dechant and Sasa \cite{dechant2020fluctuation} developed the general linear and nonlinear FRI for jump-diffusion systems on the trajectory level, Zheng and Lu \cite{zheng2024universal} found that the signal-to-noise ratio of the response of any trajectory observable to an external perturbation is upper bounded by the dynamical activity. Kwon et al. \cite{kwon2024fluctuation} also point out that the steady-state FRI directly leads to the Response Thermodynamic Uncertainty Relations (R-TUR). The use of linear algebra analysis also leads to the finding of new Fluctuation Response Relations (FRR) \cite{ptaszynski2024dissipation,aslyamov2025nonequilibrium,ptaszynski2024nonequilibrium,ptaszynski2025nonequilibrium}. Expect the interesting progress on FRIs, the use of linear algebra analysis and matrix tree theorem also provides important results on non-equilibrium response relations and inequalities \cite{harunari2024mutual,owen2020universal,owen2023size}.

In this paper, we close the gap between non-equilibrium linear response relations \cite{seifert2010fluctuation,baiesi2009nonequilibrium,baiesi2013update,maes2020response} and the FRIs \cite{dechant2020fluctuation,zheng2024universal,kwon2024fluctuation,ptaszynski2024dissipation,aslyamov2025nonequilibrium,ptaszynski2024nonequilibrium,ptaszynski2025nonequilibrium,liu2025dynamical}. We first modify the linear response relation for parameter dependent observables, and then show that the linear response relation directly leads to the FRIs. We also point out that the non-equilibrium response theory and inequalities not only apply to Markov jump processes, but also apply to general stochastic systems described by the evolution of probability distributions. Moreover, our work unravels a rich structure of the linear response relation and the FRIs. When constrained to Markov jump dynamics, the response contribution can be decomposed into correlations between dynamical events on different edges. When considering multi-parameter perturbations and multi-observable processes, our theory unifies the multi-dimensional TURs and FRIs and reveals information monotonicity properties for the multi-dimensional inequalities. At the end of this paper, we state the numerical advantage of our spatial decomposition and compare it with the traditional finite-difference method on a three-state Markov network and a large network with 100 states.

\section{Linear Response Relation for Stochastic Systems}
Throughout this paper, we consider dynamics that can be described by the probability distributions of a set of stochastic quantities $\boldsymbol{x}(t) = \{ x_1(t), x_2(t), \cdots, x_n(t) \}$. Typical examples of such systems include Markov jump processes and Langevin dynamics. The quantity $\boldsymbol{x}(t)$ could represent the position of a real physical particle or the microscopic state of a molecule in its state space. The evolution of the probability distribution $p(\boldsymbol{x}, t)$ characterizes the evolution of stochastic quantities $\boldsymbol{x}(t)$.
Another way to represent the stochastic evolution is through stochastic trajectories $X_\tau = \{ x(t) \}_{t\in[0, \tau]}$, where $\tau$ is the ultimate time of the trajectory. Consider an ensemble of the system, the evolution of each system gives a realization of the stochastic trajectory. The distribution of $\boldsymbol{x}(t)$ can be recovered from the trajectory distribution $\mathcal{P}[X_\tau]$. We further specify a trajectory observable $Q[X_\tau]$ as a functional of trajectories. The observable $Q[X_\tau]$ could either be a function of particle positions or transition events. Such observables are fundamental to many research fields, including diffusion \cite{levy1940certains,godreche2001statistics,majumdar2002local,lapolla2019manifestations,hartich2023violation}, active matter \cite{ramaswamy2010mechanics,ramaswamy2017active,marchetti2013hydrodynamics}, optics \cite{o2012time,margolin2005nonergodicity,ramesh2024arcsine,gopich2012theory}, chemical sensing \cite{bialek2005physical,endres2008accuracy,mora2010limits,govern2012fundamental,govern2014energy,harvey2023universal}, or biological transportation \cite{maffeo2012modeling,catterall2010ion,roux2004theoretical}. The average of such an observable is given by a trajectory-ensemble average:
\begin{equation}
    \langle Q \rangle = \int\mathcal{D}[X_\tau] \mathcal{P}[X_\tau]Q[X_\tau].
    \label{eq: define Q}
\end{equation}
Assuming that there is a parameter $\lambda$ controlling the dynamics of the system, such as the temperature of the thermal bath, the magnitude of the magnetic field, the strength of the energy input, or the internal energy barrier between different states. The parameter $\lambda$ generally changes the evolution operator of the system, so the distribution $p(x(t))$ and $\mathcal{P}[X_\tau]$ also change. When the parameter is changed, the observable $\langle Q \rangle$ will deviate from its original value, which gives the response of the system represented on $Q[X_\tau]$.

Historically, Agarwal \cite{agarwal1972fluctuation} developed the linear response theory around NESS. Seifert \cite{seifert2010fluctuation} and Maes \cite{baiesi2009nonequilibrium,baiesi2013update,maes2020response} have made significant progress on the stochastic trajectory level. Here, we provide a slightly different approach from the viewpoint of the trajectory score function.

The linear response of the averaged value $\langle Q \rangle$ to any parameter $\lambda$ are always defined as $\frac{\partial \langle Q \rangle}{\partial \lambda}$. It can be determined from understanding how trajectory probabilities change with the control parameter $\lambda$. This change is characterized by
\begin{equation}
    \frac{\partial \mathcal{P}[X_\tau; \lambda]}{\partial \lambda} = \Lambda[X_\tau; \lambda] \mathcal{P}[X_\tau; \lambda].
    \label{eq: traj prob response to lambda}
\end{equation}
Here we denote the control parameter's conjugate $\Lambda[X_\tau,\lambda]$ as the score function that is defined by $\Lambda \equiv {\partial  \ln \mathcal{P}[X_\tau;\lambda]}/{\partial  \lambda} $.

Given the constant normalization of trajectory probabilities, the expectation value of $\Lambda$ is always zero, $\langle \Lambda \rangle = 0$. Then by combining \cref{eq: traj prob response to lambda} and the chain rule, we arrive at a general response equality for any trajectory observable $\langle Q \rangle$:
\begin{equation}
    \frac{\partial \langle Q \rangle}{\partial \lambda} = \left\langle \frac{\partial Q}{\partial \lambda} \right\rangle + \operatorname{Cov}(Q, \Lambda), \label{eq: general response relation}
\end{equation}
Here, the covariance function is defined as $\operatorname{Cov}(Q, \Lambda) \equiv \langle Q\Lambda \rangle - \langle Q \rangle \langle \Lambda \rangle = \langle Q\Lambda \rangle$ because of $\langle \Lambda \rangle = 0$. When the observable $Q[X_{\tau}]$ does not explicitly depend on the control parameter, the first term of the r.h.s. can be ignored, and \cref{eq: general response relation} reduces to the most studied non-equilibrium linear response relations in \cite{agarwal1972fluctuation,seifert2010fluctuation,baiesi2009nonequilibrium,baiesi2013update,maes2020response}.

\subsection{Markov Jump Processes}

Throughout this paper, we take Markov jump processes as examples. However, a similar idea also applies to Langevin dynamics and more general stochastic dynamics. Now we focus on general Markov jump processes to obtain the explicit form of the linear conjugate variable $\Lambda[X_\tau]$.

The dynamics of an $n$-state Markov jump process is governed by the master equation for the time evolution of state probabilities
\begin{equation}
    \frac{\partial  \boldsymbol{p}(t)}{\partial  t} = R \cdot \boldsymbol{p}(t), \label{eq: master equation}
\end{equation}
where $\boldsymbol{p}(t) = ( p_1(t), p_2(t), \cdots, p_n(t) )^\intercal$ is the column vector of probability distributions on the $n$ states and $R = \{ R_{ij} \}_{n\times n}$ is the transition rate matrix with diagonal elements $R_{ii} = \sum_{j, j\neq i} R_{ji}$.

The stochastic trajectory $X_\tau$ of a Markov jump dynamics can be denoted by a sequence of $N$ jump events \cite{peliti2021stochastic}:
\begin{equation}
    X_\tau = ((x_0, t_0), (x_1, t_1), \cdots, (x_\alpha, t_\alpha), \cdots, (x_N, t_N)),
\end{equation}
where the system initiates at $x_0$ (at $t_0 = 0$) and undergoes a series of $N$ transitions from state $x_{\alpha-1}$ to $x_{\alpha}$ at times $t_\alpha$ with $\alpha=1,2,\cdots, N$.
For any Markov process, the trajectory probability can be generally expressed by the products of the probabilities of jump events and probabilities of dwelling in between adjacent jumps:
\begin{equation}
    \mathcal{P}[X_\tau] = p_{x_0}(t_0)\prod_{\alpha=1}^{N} R_{x_\alpha x_{\alpha-1}} \prod_{\alpha=0}^{N} e^{\int_{t_\alpha}^{t_{\alpha+1}} R_{x_\alpha x_\alpha}\mathrm{d}t}, \label{eq: traj prob}
\end{equation}
where $t_{N+1} = \tau$ denote the ultimate time of the trajectory.

It might be challenging to utilize the response relation \cref{eq: general response relation} due to the difficulties in obtaining the score function $\Lambda$ for arbitrarily controlled systems and various choices of $\lambda$. However, we demonstrate that, via temporal-spatial decomposition of the dynamics, the response can be described by the combination of the covariance between $Q$ and edge-wise local observables $\{\Theta_{ij}\}$. We first choose $\lambda = \ln R_{ij}$, the conjugate variable $\Lambda$ becomes the {\it dynamical discrepancy}, 
\begin{equation}
    \Theta_{ij}[X_\tau] \equiv N_{ij}[X_\tau] - R_{ij}T_{j}[X_\tau]
    \label{eq: Theta define}
\end{equation}
which is a fundamental quantity of our response theory. Similar quantities are also found in \cite{seifert2010fluctuation}. Here, $N_{ij}[X_\tau]$ is the total number of transitions from state $j$ to state $i$, and $T_j[X_\tau]$ is the total dwelling time on state $j$, both evaluated from a realization of the trajectory $X_\tau$. 
Averaged over an ensemble of trajectories, the expectation values of $N_{ij}$ and $R_{ij}T_j$ must equal each other for an arbitrary Markov process, and thus $\langle \Theta_{ij} \rangle=0$. For each realization of the stochastic process $X_{\tau}$, the dynamical discrepancy $\Theta_{ij}[X_\tau]$ quantifies the stochastic mismatch between the number of jumps $N_{ij}$ and the transition-rate-weighted dwelling time $R_{ij}T_j$ that is obtained by one realization (i.e., one trajectory $X_{\tau}$). The dynamical discrepancies on all transition edges $\{\Theta_{ij}\}$ offers a spatial (edge-wise) decomposition of $\Lambda$:
\begin{equation}
    \Lambda[X_\tau] = \sum_{i\neq j} \frac{\partial  \ln R_{ij}}{\partial  \lambda} \Theta_{ij}.
    \label{eq: time-independent Lambda}
\end{equation}
Therefore, the response of any trajectory observable $\langle Q \rangle$ can be decomposed into contributions from each transition edge:
\begin{equation}
    \frac{\partial  \langle Q \rangle}{\partial  \lambda} = \left\langle \frac{\partial  Q}{\partial  \lambda} \right\rangle + \sum_{i \neq j} \frac{\partial  \ln R_{ij}}{\partial \lambda} \operatorname{Cov}(Q, \Theta_{ij}).
    \label{eq: time-independent response}
\end{equation}
The above decomposition implies that a system's non-equilibrium response could be obtained by a weighted summation over correlations between the observable and the dynamical discrepancy on each transition edge. It provides both theoretical insights and numerical advantages in understanding and computing non-equilibrium system's responses to external perturbations.    Notice that the above results, \cref{eq: Theta define,eq: time-independent Lambda,eq: time-independent response}, generally hold for any time-homogeneous Markovian jump dynamics, including both NESS and non-equilibrium relaxation processes.

For time-inhomogeneous processes, where time-dependence on the transition rates $\{R_{ij}\}_{n\times n}$ is allowed in both the unperturbed transition rates $\left. R_{ij} \right|_{\lambda = 0}$ an in the perturbations $\left. 
\frac{\partial \ln R_{ij}}{\partial \lambda} \right|_{\lambda = 0}$, we obtain the most general non-equilibrium response relation with the temporal-spatial decomposition. Here, for each transition edge, we define the dynamical discrepancy accumulation rate 
\begin{equation}
    \dot \Theta_{ij}[X_{\tau}] = \dot N_{ij}(t) - R_{ij}(t)\delta_{x(t),j}
    \label{eq: Theta modifid}
\end{equation}
where $N_{ij}(t)$ is the accumulated transitions from $j$ to $i$ within time $[0,t)$ for a realization $X_\tau$, and $\delta_{x(t),j}$ is a Kronecker delta that equals unity when the system state is $x(t)=j$. Under this decomposition, the control-conjugate variable within the general non-equilibrium response relation (\cref{eq: general response relation}) can be represented by the following temporal-spatial decomposition:
\begin{equation}
    \Lambda[X_\tau] = \sum_{i \neq j} \int_0^\tau \frac{\partial \ln R_{ij}(t)}{\partial \lambda} \dot{\Theta}_{ij}[X_\tau] \mathrm{d}t.
    \label{eq: time-dependent Lambda}
\end{equation}
If both $\left. R_{ij} \right|_{\lambda = 0}$ and $\left. 
\frac{\partial \ln R_{ij}}{\partial \lambda} \right|_{\lambda = 0}$ are time-independent, \cref{eq: time-dependent Lambda} reduces to \cref{eq: time-independent Lambda}.



\section{Spatial Correlation Analysis}

The above analysis indicates that the dynamical discrepancy $\{ \Theta_{ij} \}$ is a key stochastic quantity that encodes the system information on each edge. It also has good statistical properties. Firstly, its zero-mean property is a consequence of $\langle N_{ij} \rangle = R_{ij}\langle T_j \rangle$. Secondly, the spatial correlations of $\{ \Theta_{ij} \}$ between any pair of transition edges are:
\begin{equation}
    \operatorname{Cov}(\Theta_{ij}, \Theta_{kl}) = \langle N_{ij} \rangle \delta_{(i, j), (k, l)},
    \label{eq: spatial correlation of Theta}
\end{equation}
where $\delta_{(i, j), (k, l)}$ is the Kronecker delta as defined bellow:
\begin{equation}
\delta_{(i, j), (k, l)} = 
    \begin{cases}
        1, & i = k \text{ and } j = l, \\
        0, & \text{else}.
    \end{cases}
\end{equation}
This result indicates that the dynamical discrepancies of two arbitrary edges are statistically uncorrelated. As a result, $\{ \Theta_{ij} \}$ forms a suitable basis for describing a system's response properties. \cref{eq: spatial correlation of Theta} can be proven by taking derivatives twice on both sides of the trajectory probability in \cref{eq: traj prob}. Also, \cref{eq: spatial correlation of Theta} indicates that the variance of dynamical discrepancy for each edge characterizes its average transition frequency. 

The above analysis covers the spatial decomposition of the conjugate variable $\Lambda$ into dynamical discrepancies for each edge. In the following, we further state that the decomposition of observables $Q$ implies that spatial correlations determine their non-equilibrium response properties.

Physical observables of a Markov system can usually be decomposed into ``local contributions''. In general, one can construct an observable by combining two observable types: counting and dwelling. For example, current and traffic \cite{maes2020frenesy} are transition-counting observables and state average observables are state-dwelling observables. Thus, for a general observable, we can usually decompose it as
\begin{equation}
    Q[X_\tau] = \sum_{i\neq j} a_{ij}(\lambda) \cdot N_{ij}[X_\tau] + \sum_k b_k(\lambda) \cdot T_k[X_\tau],
\end{equation}
where $a_{ij}$ is the accumulation weight associated with each transition event and $b_k$ is the weight associated with dwelling at state $k$. With the decompositions of $Q$ and $\Lambda$, one can conclude that, for any observable, the response can always be decomposed into the linear combination of spatial correlations $\operatorname{Cov}(N_{ij}, N_{kl})$, $\operatorname{Cov}(N_{ij}, T_k)$, and $\operatorname{Cov}(T_i, T_j)$ on different edges or states. Moreover, if all the coefficients $\{a_{ij}, b_k\}$ are independent of the parameter $\lambda$, the first term on the right-hand side of \cref{eq: general response relation} vanishes. In this scenario, the response of any non-equilibrium process in terms of any arbitrary observable is entirely encoded by the spatial correlations of the dynamics from the unperturbed process (i.e., transitions and dwells), which are easy to obtain. 

A particularly insightful result emerges when we consider spatially localized quantities. When both the observable $Q$ and the perturbation $\lambda$ are confined to specific edges of the network, our theory reveals an interesting property: the response depends only on the correlation between these edges, regardless of their separation in the network. For instance, by choosing $Q = a_{ij} N_{ij}$ and $\lambda = \ln R_{kl}$, where the edges $(ij)$ and $(kl)$ are arbitrarily far apart, the response $\partial \langle Q \rangle / \partial  (\ln R_{kl}) = a_{ij} \operatorname{Cov}(N_{ij}, \Theta_{kl})$ is fully determined by the statistical correlation between the two edges of interest, $i \leftarrow j$ and $k \leftarrow l$, yet no explicit information about other edges of the graph is involved.

\section{From Linear Response Equality to Fluctuation-Response Inequalities}

Our non-equilibrium response theory advances the understanding of existing response inequalities in three fundamental ways. First, by expressing Fisher information through edge-wise contributions, our theory extends Response Uncertainty Relations (RUR) to all generic types of non-equilibrium processes. Second, the spatial decomposition described in this Letter reveals deep connections to the high-dimensional Cram\'er-Rao bounds with multi-dimension observables and parameters. Third, we discover fundamental information monotonicity properties of the responsiveness inequality tightness under different choices of control parameters.

\subsection{Trajectory Fisher Information and FRI}
The theory proposed in this paper connects to existing inequalities \cite{ptaszynski2024dissipation,aslyamov2024nonequilibrium,kwon2024fluctuation,ptaszynski2024nonequilibrium}, known as RUR, via Fisher information while generalizing them from NESS to GNEP. In any generic non-equilibrium process, the Fisher information $\mathcal{I}(\lambda)=\operatorname{Var}[\Lambda]$ equals the variance of the score function $\Lambda$, which we obtain by taking the ensemble average of the derivative of \cref{eq: traj prob response to lambda}: 
\begin{subequations}
\begin{align}
    \mathcal{I}(\lambda)&=\operatorname{Var}[\Lambda] = -\left\langle \frac{\partial \Lambda}{\partial \lambda} \right\rangle \label{eq: variance Lambda} \\
    &= \left\langle \sum_{i \neq j}\int _0^\tau \left( \frac{\partial \ln R_{ij}(t)}{\partial \lambda} \right)^2 R_{ij} \delta_{x(t),j} ~{\rm d} t \right\rangle
    \label{eq: trajectory FI avg},
\end{align}
\end{subequations}
This formulation reveals how Fisher information decomposes into contributions from local dynamical events across the network. For processes with a time-independent rate matrix, it is reduced to a linear combination of averaged transitions $\mathcal{I}(\lambda) = \sum_{i\neq j} \left( \frac{\partial \ln R_{ij}}{\partial \lambda} \right)^2 \langle N_{ij} \rangle$. By choosing $\lambda=R_{ij}$, it reduces to the result in \cite{zheng2024universal}: $\mathcal{I} = \langle N_{ij} \rangle / R_{ij}^2$. Furthermore, our theory refines the result in \cite{dechant2020fluctuation} with Fisher information $\mathcal{I}(\lambda) = \langle Z_{ij}^2 R_{ij} T_j \rangle$ given the transition rates following $R_{ij} = k_{ij}e^{\lambda Z_{ij}}$.

With this expression for Fisher information, the Fluctuation-Response Inequality (FRI) \cite{dechant2020fluctuation,zheng2024universal} can be obtained from our general response relation through the Cauchy-Schwarz inequality:
\begin{equation}
    \frac{\left( \frac{\partial \langle Q \rangle}{\partial \lambda} - \left\langle \frac{\partial Q}{\partial \lambda} \right\rangle \right)^2}{\operatorname{Var}[Q]} \le \operatorname{Var}[\Lambda] = \mathcal{I}(\lambda).
    \label{eq: general CR bound}
\end{equation}
Furthermore, our theory indicates that the FRI bounds saturate if and only if the observable $Q_s$ and the score function $\Lambda$ are linearly dependent:
\begin{equation} 
    Q_s \propto \Lambda = \sum_{i\neq j} \frac{\partial \ln R_{ij}}{\partial \lambda} N_{ij} + \sum_{k}R_{kk}\frac{\partial \ln R_{kk}}{\partial \lambda}T_{k}.
\end{equation}

\subsection{Multi-dimensional Cram\'er-Rao Inequality}
Our framework reveals a rich structure for processes with multi-dimensional observables or control parameters through high-dimensional Cram\'er-Rao inequalities.
There are two types of multi-dimensional Cram\'er-Rao inequalities:
\begin{itemize}
    \item[(1)] The first type addresses multiple observables. For $K$ observables $\boldsymbol{Q}^{(K)} = (Q_1, \cdots, Q_K)$, one obtains
    \begin{equation}
        (\partial_\lambda\langle \boldsymbol{Q}^{(K)} \rangle)^\intercal (\Xi^{(K)}_{Q})^{-1}(\partial_\lambda\langle \boldsymbol{Q}^{(K)} \rangle) \le \mathcal{I}(\lambda),
    \end{equation} 
    where $(\Xi^{(K)}_{Q})_{ij} = \operatorname{Cov}(Q_i, Q_j)$ is the covariance matrix of $\boldsymbol{Q}^{(K)}$. This result leads to multi-dimensional Thermodynamic Uncertainty Relations (TUR) and FRI as reported in \cite{dechant2018multidimensional,dechant2021improving}.
    \item[(2)] The second type involves multiple parameters $\lambda \in \mathbb{R}^K$, yielding:
    \begin{equation}
        \Psi^{\lambda} \equiv (\partial_{\lambda}\langle Q \rangle)^\intercal (\mathcal{I}(\lambda))^{-1}(\partial_{\lambda}\langle Q \rangle) \le \operatorname{Var}[Q],
    \end{equation}
    where $\partial_{\lambda}\langle Q \rangle = (\partial_{\lambda_1}\langle Q \rangle, \cdots, \partial_{\lambda_K}\langle Q \rangle)$ and the entries of Fisher information matrix is $(\mathcal{I}(\lambda))_{ij} = -\langle \partial_{\lambda_i}\partial_{\lambda_j}\ln\mathcal{P}[X_\tau]\rangle$. Recent work \cite{kwon2024fluctuation} reported that the multi-parameter inequality leads to RUR and the recent conjectured R-TUR \cite{ptaszynski2024dissipation}.
\end{itemize}
When the observable $Q$ is $\lambda$-dependent, the $(\partial_{\lambda}\langle Q \rangle)$ should be replaced by $(\partial_\lambda \langle Q \rangle - \langle \partial_\lambda Q \rangle)$ for the above inequalities. Interestingly, we can prove that the two types of multi-dimensional Cram\'er-Rao inequalities are equivalent when we choose dynamical discrepancies as observables $Q_\alpha = \Lambda_\alpha$ for all $2 \le \alpha \le K$. Detailed derivations are discussed in the supplementary material.

The spatial correlations of $\{\Theta_{ij}\}$ in \cref{eq: spatial correlation of Theta} implies that the multi-parameter Fisher information matrix is diagonal if and only if each parameter only affects one local edge. 
Consider a more general scenario, where each parameter $\lambda_\alpha$ affects $n_\alpha$ edges $\boldsymbol{e}^{(\alpha)} = (e^{(\alpha)}_1, e^{(\alpha)}_2, \cdots, e^{(\alpha)}_{n_\alpha})$, where $\alpha \in (1, 2, \cdots, K)$. If the controls are mutually exclusive: $\boldsymbol{e}^{(\alpha)} \cap \boldsymbol{e}^{(\alpha')} = \emptyset$ for any $\alpha \neq \alpha'$, then the multi-parameter Fisher information becomes diagonal, significantly simplifying the bounds.

Our analysis of the Cram\'er-Rao inequality applies to arbitrarily non-equilibrium dynamics, including time-dependent processes. Here, we find out that the RUR for generic non-equilibrium processes assumes the same form as the previously reported RUR for NESS \cite{kwon2024fluctuation}. To illustrate this, consider localized control parameters as edge kinetic barrier $B_{ij} = B_{ji}$ or edge force $F_{ij} = - F_{ji}$ from the rate formula
\begin{equation}
    R_{ij} = e^{B_{ij} + F_{ij}/2},
\end{equation}
where each parameter only affects a single edge. In this case, our inequality extends the following two sets of RURs
\begin{align}
    \sum_{i \le j} \frac{(\partial_{B_{ij}} \langle Q \rangle)^2}{\langle N_{ij} \rangle} \le \operatorname{Var}[Q], \quad 
    \sum_{i \le j} \frac{4(\partial_{F_{ij}} \langle Q \rangle)^2}{\langle N_{ij} \rangle} \le \operatorname{Var}[Q],
\end{align}
from NESS \cite{kwon2024fluctuation} to arbitrary time-dependent processes.

Ultimately, these multi-dimension Cram\'er-Rao analyses reveal two fundamental monotonicity properties that characterize the information content of response inequalities.
First, adding extra parameters leads to tighter bounds. The information monotonicity states that $\Psi^{\lambda} \le \Psi^{\lambda, \lambda'}$. The inequality saturates if and only if the newly added parameter linearly depends on the old ones. The proof of the inequality is similar to the one in \cite{dechant2020fluctuation}. 
Second, separating a global perturbation into independent local ones leads to tighter bounds. For a global parameter $\lambda^*$ that affects $\boldsymbol{e}^{(*)}$ edges, its effect can be represented by some local parameters as $\lambda^* = \sum_{\alpha}\frac{\partial \lambda^*}{\partial \lambda_\alpha}\lambda_\alpha$ with $\boldsymbol{e}^{(\alpha)} \cap \boldsymbol{e}^{(\alpha')} = \emptyset$ and $\bigcup_{\alpha}\boldsymbol{e}^{(\alpha)} = \boldsymbol{e}^{(*)}$. In this case, the information monotonicity reads
\begin{equation}
    \frac{(\partial_{\lambda^*} \langle Q \rangle)^2}{\operatorname{Var}[\Lambda^*]} \le \sum_\alpha \frac{(\partial_{\lambda_\alpha} \langle Q \rangle)^2}{\operatorname{Var}[\Lambda_\alpha]} \le \operatorname{Var}[Q].
\end{equation}

\section{Numerical Efficiency}

Our response equality leads to a novel numerical approach, significantly reducing the sampling cost for computing system responses. Traditional estimation of the system's response to external control parameters typically involves finite-difference methods that suffer from an inherent accuracy-cost trade-off. Furthermore, the numerical effort to estimate a multi-dimensional control response gradient $\lambda = (\lambda_1, \lambda_2, \cdots, \lambda_{N_\text{para}})$ scales with the number of parameters. Our approach allows us to efficiently estimate multi-variable response properties by numerically sampling trajectories from a single unperturbed simulation. 

When estimating a system's response to a given control parameter $\lambda$, the traditional finite-difference method estimates responses by performing two sets of simulations and comparing the differences:
\begin{equation}
    \frac{\partial \langle Q \rangle}{\partial  \lambda} \approx \eta \equiv \frac{\langle Q \rangle_{\text{ptb}} - \langle Q \rangle_{\text{unptb}}}{\Delta \lambda}.
\end{equation}
where the expectation $\langle Q \rangle$ is estimated from an ensemble of $N_{\text{traj}}$ independent stochastic trajectories which can be obtained by repeated kinetic Monte Carlo simulations:
\begin{equation}
    \langle Q \rangle
    \approx \overline{Q}
    \equiv \frac{1}{N_{\text{traj}}} \sum_{i = 1}^{N_{\text{traj}}} Q[X_\tau^{(i)}],
\end{equation}
where $X_\tau^{(i)}$ stands for the $i$-th sampled trajectory.

Firstly, the finite-difference approach suppers from {\it accuracy--precision--cost trade-off}: On the one hand, unless the response is strictly linear, the estimated response sensitivity carries a systematic error for any finite $\Delta \lambda$, creating a trade-off between accuracy and precision at a given computational cost. On the other hand, there is a trade-off between the precision and the computational cost: since the difference in observable decays with $\Delta \lambda$, resulting in the need for a larger sampling size $N_{\rm traj}$. According to the law of large numbers, $\overline{Q} \to \mathcal{N}(\mu(Q), \frac{\sigma^2(Q)}{N_{\text{traj}}})$ as $N_\text{traj} \to +\infty$ with $\mu$ and $\sigma$ representing expectation and standard deviation, respectively. The finite-difference resolution between the perturbed and unperturbed observables can be captured by the signal-to-noise ratio:
\begin{subequations}
\begin{align}
    \text{SNR} &\equiv \frac{|\mu_{\text{ptb}}(\overline{Q}) - \mu_{\text{unptb}}(\overline{Q})|}{\sqrt{\frac{\sigma^2_\text{ptb}(\overline{Q}) + \sigma^2_\text{unptb}(\overline{Q})}{2}}} \\
    &\approx \frac{|\eta \Delta\lambda| \cdot \sqrt{2 N_\text{traj}}}{\sqrt{\operatorname{Var}_\text{ptb}[Q] + \operatorname{Var}_\text{unptb}[Q]}}.
\end{align}
\end{subequations}
This analysis illustrated the precision-cost trade-off relation that the required sampling size scales with $N_{\text{traj}} \propto \frac{1}{\Delta \lambda^2}$. Therefore, to achieve higher accuracy (low systematic error), one needs a smaller $\Delta \lambda$, which results in a larger number of sampled trajectories to maintain the same resolution. 

Secondly, traditional finite-difference requires a large number of simulations to estimate the sensitivity gradient of a high-dimensional control parameter vector $\lambda$. In this case, the same finite-difference procedure needs to be repeated $N_\text{para}$ times to obtain the response properties for all parameters. 

In contrast, our response equality \cref{eq: general response relation} avoids both difficulties by converting the sensitivity estimation problem into an estimation of correlations of different dynamical events. This correlation estimation does not suffer from the accuracy-sampling trade-off difficulty illustrated above, and it only requires one set of simulations (on the unperturbed system) to obtain all partial derivatives within the sensitivity gradient.

\subsection{Numerical Comparison on a Three-state Markov Network}

Our universal response equality \cref{eq: general response relation} together with the dynamical discrepancy \cref{eq: Theta define,eq: Theta modifid} significantly reduces the computational cost for response properties. 

\begin{figure}[htbp]
    \centering
    \includegraphics[width=1\linewidth]{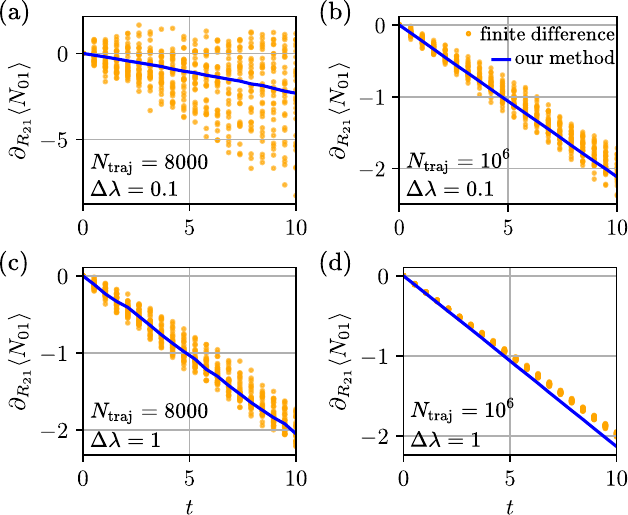}
    \caption{Four sets of kinetic Monte Carlo simulations on a three-state Markov system with the rate matrix in \cref{eq: rate matrix}. Each point and curve is obtained from trajectory averages. The traditional finite-difference method is repeated 30 times for each subfigure to show its convergence issue. The number of trajectories $N_\text{traj}$ is $8000$ for (a) and (c), and is $10^6$ for (b) and (d). The $\Delta \lambda$ in the finite-difference method is $0.1$ for (a) and (b), and is $1$ for (c) and (d).}
    \label{fig: three state response}
\end{figure}

We first demonstrate the numerical advantages of our method over the finite-difference method on a three-state Markov system. Consider the following unperturbed rate matrix
\begin{equation}
    R = \begin{pmatrix}
        -130 & 30 & 70 \\
        90 & -40 & 50 \\
        40 & 10 & -120
        \label{eq: rate matrix}
    \end{pmatrix}
\end{equation}
with the system initially prepared at state 1. We define the observable as the accumulated number of transitions from state 1 to state 0. To estimate the response of $\langle N_{01} \rangle$ to the control $\lambda = R_{21}$, we compare our method with the finite-difference method, as shown in \cref{fig: three state response}.

We have mentioned that the traditional finite-difference method suffers from an accuracy-precision-cost trade-off relation. This issue is clearly illustrated by the numerical results represented by the orange dots in \cref{fig: three state response}. Comparing \cref{fig: three state response}(a) with (b) (or (c) with (d)), it shows that the larger the sampling size, the higher the precision (less spread of orange dots). The deviation between the orange dots and the blue curve (our method) in \cref{fig: three state response}(d) shows that a finite $\Delta \lambda$ leads to a systematic error. A smaller $\Delta \lambda$ may improve this issue, but it immediately gives rise to a convergence issue as shown by comparing \cref{fig: three state response}(b) with (d) (or (a) with (c)). In contrast, our method is free from this trade-off issue. The four blue curves in \cref{fig: three state response} show a consistent trend, representing the high accuracy, high precision, and low cost of our method. Also, by comparing our result with the finite-difference result shown in \cref{fig: three state response}(b), it indicates that the precision of the finite-difference method is worse than ours, even when we choose a relatively large sample size. \cref{fig: three state response}(d) indicates that the systematic error of the finite-difference method is more prominent for the large finite $\Delta \lambda$. 

\subsection{Numerical Comparison on a Large Markov Network}

\begin{figure}[htbp]
    \centering
    \includegraphics[width=1\linewidth]{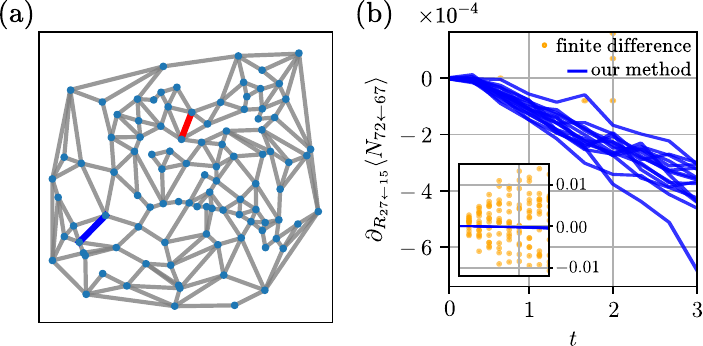}
    \caption{A Markov network graph with $100$ states, with its transition rates randomly generated with values between $10$ and $100$. This system's kinetic Monte Carlo simulations are performed with the same initial probability (randomly generated).  (a) The input edge (control) is in red, and the output edge (observable) is in blue. (b) In each realization, we perform the finite-difference method (orange dots) and our method (blue curves) for the sampling size $N_{\text{traj}} = 2\times 10^{5}$. We repeat these realizations $15$ times to illustrate the variances. The parameter difference for the traditional finite-difference method is $\Delta \lambda = 1$. }
    \label{fig: large network}
\end{figure}

In the following example, we numerically illustrate that the advantage of our method is more prominent in larger networks. Here, we randomly generate a Markov graph with $100$ states, as shown in \cref{fig: large network}(a). Starting from a non-stationary initial probability distribution, the system evolves toward the steady state corresponding to the rate matrix of the graph. In \cref{fig: large network}(b), we compare the sensitivity results from traditional finite-difference approaches (orange dots) with those from our method (blue curves) under the same sampling size. By repeating the simulations $15$ times, our results indicate that our method captures the system's sensitivity well, yet the traditional finite-difference method suffers from both low precision (extremely large variance) and low accuracy (positive systematic error).

\section{Conclusion}
This work finds a novel spatial decomposition for the linear response properties of Markov jump dynamics. By decomposing the linear response conjugate variable, we decompose the linear response property of any trajectory observable into contributions of dynamical events on different local edges. This decomposition naturally resolves the accuracy-precision-cost trade-off of the traditional finite-difference method for response simulations, enabling efficient numerical evaluation of response properties from the unperturbed trajectory data. Our work also fills the gap between non-equilibrium linear response relations and the recently developed FRIs. By using information theory, we find the connections between the two on the trajectory level. This unification also unravels rich structures of the multi-parameter FRIs, which leads to the unification of multi-parameter FRIs and multi-observable TURs.  This work also reveals fundamental monotonicity principles regarding the information encoded in multiple degrees of freedom of control parameters. These advances provide theoretical insights and practical tools for studying complex networks and biological systems arbitrarily far from equilibrium.


\section*{Acknowledgements}
This work is supported by the National Science Foundation under Grant No. DMR-2145256 and Alfred P. Sloan Foundation Award under grant number G-2025-25194.

\appendix
\section{Equivalence between two types of multi-dimensional Cram\'er-Rao inequalities} \label{SIsec: equivalence}

Below we sketch the derivation of the equivalence between the two types of multi-dimensional Cram\'er-Rao inequalities as discussed in the main text.

We start the derivation by introducing the following lemma \cite{bhatia2013matrix}.

\begin{lemma}(Block Matrix Inverse) \label{SIlm: block matrix inverse}
    Given a block matrix
    \begin{equation}
        \Phi \equiv \begin{pmatrix}
            A & B \\
            C & D
        \end{pmatrix},
    \end{equation}
    if $D$ is invertible, define the Schur complement of $D$ as
    \begin{equation}
        \Sigma \equiv A - B D^{-1} C.
    \end{equation}
    $\Phi$ is invertible if and only if $\Sigma$ is invertible. Its inverse is
    \begin{equation}
        \Phi^{-1} = \begin{pmatrix}
            \Sigma^{-1} & -\Sigma^{-1} B D^{-1} \\
            -D^{-1} C \Sigma^{-1} & D^{-1} C \Sigma^{-1} B D^{-1} + D^{-1}
        \end{pmatrix}.
    \end{equation}
\end{lemma}

The multi-observable Cram\'er-Rao inequality reads as
\begin{widetext}
\begin{equation}
    \begin{pmatrix}
        \partial_\lambda \langle Q_1 \rangle - \langle \partial_\lambda Q_1 \rangle \\ \partial_\lambda \langle Q_2 \rangle - \langle \partial_\lambda Q_2 \rangle \\ \vdots \\ \partial_\lambda \langle Q_K \rangle - \langle \partial_\lambda Q_K \rangle
    \end{pmatrix}^{\intercal}
    \begin{pmatrix}
        \operatorname{Var}[Q_1] & \operatorname{Cov}(Q_1, Q_2) & \cdots & \operatorname{Cov}(Q_1, Q_K) \\
        \operatorname{Cov}(Q_1, Q_2) & \operatorname{Var}[Q_2] & \cdots & \operatorname{Cov}(Q_2, Q_K) \\
        \vdots & \vdots & \ddots & \vdots \\
        \operatorname{Cov}(Q_1, Q_K) & \operatorname{Cov}(Q_2, Q_K) & \cdots & \operatorname{Var}[Q_K]
    \end{pmatrix}^{-1}
    \begin{pmatrix}
        \partial_\lambda \langle Q_1 \rangle - \langle \partial_\lambda Q_1 \rangle \\ \partial_\lambda \langle Q_2 \rangle - \langle \partial_\lambda Q_2 \rangle \\ \vdots \\ \partial_\lambda \langle Q_K \rangle - \langle \partial_\lambda Q_K \rangle
    \end{pmatrix}
    \le \operatorname{Var}[\Lambda].
    \label{SIeq: multi-observable CR bound}
\end{equation}
\end{widetext}

Now we choose $Q_\alpha = \Lambda_\alpha$ for all $2 \le \alpha \le K$. Notice that in general $\partial_{\lambda_\alpha}\langle \Lambda_{\alpha'} \rangle - \left\langle \partial_{\lambda_\alpha}\Lambda_{\alpha'} \right\rangle = \operatorname{Cov}(\Lambda_\alpha, \Lambda_{\alpha'}) \neq 0$. Without loss of generality, we rename $Q_1$, $\lambda$, and $\Lambda$ as $Q$, $\lambda_1$, and $\Lambda_1$, respectively. In this case, \cref{SIeq: multi-observable CR bound} becomes
\begin{widetext}
\begin{equation}
    \begin{pmatrix}
        \partial_{\lambda_1} \langle Q \rangle - \langle \partial_{\lambda_1} Q \rangle \\ \operatorname{Cov}(\Lambda_1, \Lambda_2) \\ \vdots \\ \operatorname{Cov}(\Lambda_1, \Lambda_K)
    \end{pmatrix}^{\intercal}
    \begin{pmatrix}
        \operatorname{Var}[Q] & \operatorname{Cov}(Q, \Lambda_2) & \cdots & \operatorname{Cov}(Q, \Lambda_K) \\
        \operatorname{Cov}(Q, \Lambda_2) & \operatorname{Var}[\Lambda_2] & \cdots & \operatorname{Cov}(\Lambda_2, \Lambda_K) \\
        \vdots & \vdots & \ddots & \vdots \\
        \operatorname{Cov}(Q, \Lambda_K) & \operatorname{Cov}(\Lambda_2, \Lambda_K) & \cdots & \operatorname{Var}[\Lambda_K]
    \end{pmatrix}^{-1}
    \begin{pmatrix}
        \partial_{\lambda_1} \langle Q \rangle - \langle \partial_{\lambda_1} Q \rangle \\ \operatorname{Cov}(\Lambda_1, \Lambda_2) \\ \vdots \\ \operatorname{Cov}(\Lambda_1, \Lambda_K)
    \end{pmatrix}
    \le \operatorname{Var}[\Lambda_1].
    \label{SIeq: multi-observable CR bound Lambda}
\end{equation}
\end{widetext}
Then, we divide the matrix and vectors in \cref{SIeq: multi-observable CR bound Lambda} into blocks and rearrange it as follows:
\begin{equation}
    \frac{1}{c'}
    \begin{pmatrix}
        x \\
        a^\intercal
    \end{pmatrix}^\intercal
    \begin{pmatrix}
        c & b \\
        b^\intercal & \mathcal{I}
    \end{pmatrix}^{-1}
    \begin{pmatrix}
        x \\ a^\intercal
    \end{pmatrix}
    \le 1,
    \label{SIeq: multi-observable block}
\end{equation}
where $x = \partial_{\lambda_1} \langle Q \rangle - \langle \partial_{\lambda_1} Q \rangle$, $a = \begin{pmatrix}
    \operatorname{Cov}(\Lambda_1, \Lambda_2) & \operatorname{Cov}(\Lambda_1, \Lambda_3) & \cdots & \operatorname{Cov}(\Lambda_1, \Lambda_K)
\end{pmatrix}$, $b = \begin{pmatrix}
    \operatorname{Cov}(Q, \Lambda_2) & \operatorname{Cov}(Q, \Lambda_3) & \cdots & \operatorname{Cov}(Q, \Lambda_K)
\end{pmatrix}$, $c = \operatorname{Var}[Q]$, $c' = \operatorname{Var}[\Lambda_1]$, and $\mathcal{I}$ is the $(K-1)\times (K-1)$ Fisher information matrix
\begin{equation}
    \mathcal{I} = \begin{pmatrix}
        \operatorname{Var}[\Lambda_2] & \cdots & \operatorname{Cov}(\Lambda_2, \Lambda_K) \\
        \vdots & \ddots & \vdots \\
        \operatorname{Cov}(\Lambda_2, \Lambda_K) & \cdots & \operatorname{Var}[\Lambda_K]
    \end{pmatrix}.
\end{equation}

Now we use \cref{SIlm: block matrix inverse} by setting
\begin{equation}
    \Phi = \begin{pmatrix}
        c & b \\
        b^\intercal & \mathcal{I}
    \end{pmatrix}.
\end{equation}
Let us assume that $\Lambda_2, \cdots, \Lambda_K$ are linearly independent so that the Fisher information matrix $\mathcal{I}$ is invertible. Therefore, the Schur complement of $\mathcal{I}$ is
\begin{equation}
    \Sigma = c - b \mathcal{I}^{-1} b^\intercal.
\end{equation}
Now we check that $\Sigma > 0$ (positive definite as a $1 \times 1$ matrix). The Schur determinant formula \cite{bhatia2013matrix} states
\begin{equation}
    \det\Phi = \det\Sigma \det\mathcal{I} = (c - b \mathcal{I}^{-1} b^\intercal) \det\mathcal{I}.
\end{equation}
The matrices $\Phi$ and $\mathcal{I}$ are all positive definite since they are covariance matrices. Therefore, we have $\Sigma = c - b \mathcal{I}^{-1} b^\intercal > 0$ and so $\Sigma$ is invertible.

Plugin in the formula of $\Phi^{-1}$ immediately yields
\begin{subequations}
\begin{align}
    \Phi^{-1} &= \begin{pmatrix}
        \Sigma^{-1} & -\Sigma^{-1} b^\intercal \mathcal{I}^{-1} \\
        -\mathcal{I}^{-1} b \Sigma^{-1} & \mathcal{I}^{-1} b \Sigma^{-1} b^\intercal \mathcal{I}^{-1} + \mathcal{I}^{-1}
    \end{pmatrix} \\
    &= \Sigma^{-1} \begin{pmatrix}
        1 & -b \mathcal{I}^{-1} \\
        -\mathcal{I}^{-1} b^\intercal & \mathcal{I}^{-1} b^\intercal b \mathcal{I}^{-1}
    \end{pmatrix} + \begin{pmatrix}
        0 & 0 \\
        0 & \mathcal{I}^{-1}
    \end{pmatrix} \\
    &= \frac{1}{c - b \mathcal{I}^{-1} b^\intercal} \begin{pmatrix}
        1 & -b \mathcal{I}^{-1} \\
        -\mathcal{I}^{-1} b^\intercal & \mathcal{I}^{-1} b^\intercal b \mathcal{I}^{-1}
    \end{pmatrix} + \begin{pmatrix}
        0 & 0 \\
        0 & \mathcal{I}^{-1}
    \end{pmatrix}.
    \label{SIeq: multi-observable inverse}
\end{align}
\end{subequations}
Plugin \cref{SIeq: multi-observable inverse} into \cref{SIeq: multi-observable block} yields
\begin{equation}
    \frac{a \mathcal{I}^{-1} a^\intercal}{c'} + \frac{\begin{pmatrix}
        x \\ a
    \end{pmatrix}^\intercal \begin{pmatrix}
        1 & -b \mathcal{I}^{-1} \\
        -\mathcal{I}^{-1} b^\intercal & \mathcal{I}^{-1} b^\intercal b \mathcal{I}^{-1}
    \end{pmatrix} \begin{pmatrix}
        x \\ a
    \end{pmatrix}}{c' (c - b \mathcal{I}^{-1} b^\intercal)} \le 1.
\end{equation}
Multiplying $c'(c - b \mathcal{I}^{-1} b^\intercal)$ on both sides yields (notice that $c > 0$, $c' > 0$, $\Sigma = c - b \mathcal{I}^{-1} b^\intercal > 0$)
\begin{align}
    a \mathcal{I}^{-1} a^\intercal (c - b \mathcal{I}^{-1} b^\intercal) &+ \begin{pmatrix}
        x \\ a
    \end{pmatrix}^\intercal \begin{pmatrix}
        1 & -b \mathcal{I}^{-1} \\
        -\mathcal{I}^{-1} b^\intercal & \mathcal{I}^{-1} b^\intercal b \mathcal{I}^{-1}
    \end{pmatrix} \begin{pmatrix}
        x \\ a
    \end{pmatrix} \nonumber \\
    &\le c'(c - b \mathcal{I}^{-1} b^\intercal).
    \label{SIeq: original inequality}
\end{align}
Rearranging it yields the equivalent inequality
\begin{align}
    b \mathcal{I}^{-1} b^\intercal (c' - a \mathcal{I}^{-1} a^\intercal) &+ \begin{pmatrix}
        x \\ a
    \end{pmatrix}^\intercal \begin{pmatrix}
        1 & -b \mathcal{I}^{-1} \\
        -\mathcal{I}^{-1} b^\intercal & \mathcal{I}^{-1} b^\intercal b \mathcal{I}^{-1}
    \end{pmatrix} \begin{pmatrix}
        x \\ a
    \end{pmatrix} \nonumber \\
    &\le c(c' - a \mathcal{I}^{-1} a^\intercal).
\end{align}
We further notice that
\begin{align}
    &\begin{pmatrix}
        x \\ a
    \end{pmatrix}^\intercal \begin{pmatrix}
        1 & -b \mathcal{I}^{-1} \\
        -\mathcal{I}^{-1} b^\intercal & \mathcal{I}^{-1} b^\intercal b \mathcal{I}^{-1}
    \end{pmatrix} \begin{pmatrix}
        x \\ a
    \end{pmatrix} \nonumber \\
    ={}& \begin{pmatrix}
        x \\ b
    \end{pmatrix}^\intercal \begin{pmatrix}
        1 & -a \mathcal{I}^{-1} \\
        -\mathcal{I}^{-1} a^\intercal & \mathcal{I}^{-1} a^\intercal a \mathcal{I}^{-1}
    \end{pmatrix} \begin{pmatrix}
        x \\ b
    \end{pmatrix}.
\end{align}
It is due to $(a \mathcal{I}^{-1} b^\intercal)^\intercal = b (\mathcal{I}^{-1})^\intercal a^\intercal = b \mathcal{I}^{-1} a^\intercal$ since $\mathcal{I}^{-1}$ is symmetric. Therefore, the multi-observable type inequality is equivalent to the following inequality
\begin{align}
    b \mathcal{I}^{-1} b^\intercal (c' - a \mathcal{I}^{-1} a^\intercal) &+ \begin{pmatrix}
        x \\ b
    \end{pmatrix}^\intercal \begin{pmatrix}
        1 & -a \mathcal{I}^{-1} \\
        -\mathcal{I}^{-1} a^\intercal & \mathcal{I}^{-1} a^\intercal a \mathcal{I}^{-1}
    \end{pmatrix} \begin{pmatrix}
        x \\ b
    \end{pmatrix} \nonumber \\
    &\le c(c' - a \mathcal{I}^{-1} a^\intercal).
    \label{SIeq: rearranged inequality}
\end{align}

Comparing \cref{SIeq: rearranged inequality} with \cref{SIeq: original inequality} we can find that it swaps the positions of $c$ and $c'$ and of $a$ and $b$ simultaneously. As a consequence, the following two inequalities are equivalent:
\begin{subequations}
    \begin{align}
    \begin{pmatrix}
        x \\
        a^\intercal
    \end{pmatrix}^\intercal
    \begin{pmatrix}
        c & b \\
        b^\intercal & \mathcal{I}
    \end{pmatrix}^{-1}
    \begin{pmatrix}
        x \\ a^\intercal
    \end{pmatrix}
    &\le c', \\
    \begin{pmatrix}
        x \\
        b^\intercal
    \end{pmatrix}^\intercal
    \begin{pmatrix}
        c' & a \\
        a^\intercal & \mathcal{I}
    \end{pmatrix}^{-1}
    \begin{pmatrix}
        x \\ b^\intercal
    \end{pmatrix}
    &\le c,
\end{align}
\end{subequations}
which proved our statement that the multi-observable Cram\'er-Rao inequality and multi-parameter Cram\'er-Rao inequality are equivalent when choosing $Q_\alpha = \Lambda_\alpha$ for all $2 \le \alpha \le K$.

\section{Information Monotonicity} \label{SIsec: information monotonicity}
\subsection{Information Monotonicity for Adding Parameters} \label{SIsubsec: parameter monotonicity}

Following the notations in our main text, the left-hand side of the multi-parameter Cram\'er-Rao inequality is denoted as
\begin{widetext}
\begin{subequations}
\begin{align}
    \Psi^{\boldsymbol{\lambda}}
    &= (\partial_{\boldsymbol{\lambda}}\langle Q \rangle - \langle \partial_{\boldsymbol{\lambda}} Q \rangle)^\intercal (\mathcal{I}(\boldsymbol{\lambda}))^{-1}(\partial_{\boldsymbol{\lambda}}\langle Q \rangle - \langle \partial_{\boldsymbol{\lambda}} Q \rangle) \\
    &= \begin{pmatrix}
        \partial_{\lambda_1} \langle Q \rangle - \langle \partial_{\lambda_1} Q \rangle \\ \partial_{\lambda_2} \langle Q \rangle - \langle \partial_{\lambda_2} Q \rangle \\ \vdots \\ \partial_{\lambda_K} \langle Q \rangle - \langle \partial_{\lambda_K} Q \rangle
    \end{pmatrix}^{\intercal}
    \begin{pmatrix}
        \operatorname{Var}[\Lambda_1] & \operatorname{Cov}(\Lambda_1, \Lambda_2) & \cdots & \operatorname{Cov}(\Lambda_1, \Lambda_K) \\
        \operatorname{Cov}(\Lambda_1, \Lambda_2) & \operatorname{Var}[\Lambda_2] & \cdots & \operatorname{Cov}(\Lambda_2, \Lambda_K) \\
        \vdots & \vdots & \ddots & \vdots \\
        \operatorname{Cov}(\Lambda_1, \Lambda_K) & \operatorname{Cov}(\Lambda_2, \Lambda_K) & \cdots & \operatorname{Var}[\Lambda_K]
    \end{pmatrix}^{-1}
    \begin{pmatrix}
        \partial_{\lambda_1} \langle Q \rangle - \langle \partial_{\lambda_1} Q \rangle \\ \partial_{\lambda_2} \langle Q \rangle - \langle \partial_{\lambda_2} Q \rangle \\ \vdots \\ \partial_{\lambda_K} \langle Q \rangle - \langle \partial_{\lambda_K} Q \rangle
    \end{pmatrix}.
\end{align}
\end{subequations}
\end{widetext}
For simplicity, we denote the vector $(\partial_{\boldsymbol{\lambda}}\langle Q \rangle - \langle \partial_{\boldsymbol{\lambda}} Q \rangle)$ as $\boldsymbol{v}^{(K)}$. We also use the following block matrix notations:
\begin{equation}
    \mathcal{I}(\boldsymbol{\lambda}, \lambda') \equiv \begin{pmatrix}
        \mathcal{I}' & m \\
        m^\intercal & \boldsymbol{\mathcal{I}}
    \end{pmatrix},
\end{equation}
where $\mathcal{I}' \equiv \mathcal{I}(\lambda')$, $\boldsymbol{\mathcal{I}} = \mathcal{I}(\boldsymbol{\lambda})$, and $m = \begin{pmatrix}
    \operatorname{Cov}(\Lambda', \Lambda_1) & \operatorname{Cov}(\Lambda', \Lambda_2) & \cdots & \operatorname{Cov}(\Lambda', \Lambda_K)
\end{pmatrix}$.

For $K+1$ parameters, we use \cref{SIlm: block matrix inverse} by setting $\Phi = \mathcal{I}(\boldsymbol{\lambda}, \lambda')$, $A = \mathcal{I}(\lambda')$, and $D = \mathcal{I}(\boldsymbol{\lambda})$. In this case, we have
\begin{subequations}
\begin{align}
    \Phi^{-1} ={}& \frac{1}{\operatorname{Var}[\Lambda'] - m \boldsymbol{\mathcal{I}}^{-1} m^\intercal} \begin{pmatrix}
        1 & -m \boldsymbol{\mathcal{I}}^{-1} \\
        -\boldsymbol{\mathcal{I}}^{-1} m^\intercal & \boldsymbol{\mathcal{I}}^{-1} m^\intercal m \boldsymbol{\mathcal{I}}^{-1}
    \end{pmatrix} \nonumber \\
    &+ \begin{pmatrix}
        0 & 0 \\
        0 & \boldsymbol{\mathcal{I}}^{-1}
    \end{pmatrix} \\
    ={}& \frac{1}{\Sigma}M + \begin{pmatrix}
        0 & 0 \\
        0 & \boldsymbol{\mathcal{I}}^{-1}
    \end{pmatrix},
\end{align}
\end{subequations}
where $\Sigma = \operatorname{Var}[\Lambda'] - m \boldsymbol{\mathcal{I}}^{-1} m^\intercal$ and $M = \begin{pmatrix}
        1 & -m \boldsymbol{\mathcal{I}}^{-1} \\
        -\boldsymbol{\mathcal{I}}^{-1} m^\intercal & \boldsymbol{\mathcal{I}}^{-1} m^\intercal m \boldsymbol{\mathcal{I}}^{-1}
\end{pmatrix}$. Similar to the last section, we can prove that $\Sigma > 0$ and the matrix $M$ is positive semi-definite. Therefore, we have
\begin{subequations}
\begin{align}
    \Psi^{\boldsymbol{\lambda}, \lambda'} &= {\boldsymbol{v}^{(K+1)}}^\intercal (\mathcal{I}(\boldsymbol{\lambda}, \lambda'))^{-1} \boldsymbol{v}^{(K+1)} \\
    &= \frac{1}{\Sigma} {\boldsymbol{v}^{(K+1)}}^\intercal M \boldsymbol{v}^{(K+1)} + {\boldsymbol{v}^{(K)}}^\intercal (\mathcal{I}(\boldsymbol{\lambda}))^{-1} \boldsymbol{v}^{(K)} \\
    &\ge {\boldsymbol{v}^{(K)}}^\intercal (\mathcal{I}(\boldsymbol{\lambda}))^{-1} \boldsymbol{v}^{(K)} \\
    &= \Psi^{\boldsymbol{\lambda}},
\end{align}
\end{subequations}
where $\boldsymbol{v}^{(K)}$ is the vector $\boldsymbol{v}^{(K+1)}$ with the component correspond to $\lambda'$ removed.

\subsection{Information Monotonicity for Global Parameter Decomposition} \label{SIsubsec: decomposition monotonicity}

For a global parameter $\lambda^*$ that affects $\boldsymbol{e}^{(*)}$ edges, its effect can be represented by some local parameters $\lambda^* = \sum_{\alpha} \frac{\partial \lambda^*}{\partial \lambda_\alpha} \lambda_\alpha$. We further assume that $\boldsymbol{e}^{(\alpha)} \cap \boldsymbol{e}^{(\alpha')} = \emptyset$ and $\bigcup_{\alpha}\boldsymbol{e}^{(\alpha)} = \boldsymbol{e}^{(*)}$, so that they are independent of each other.

For simplicity, we discuss time-independent systems here. Similar derivations also apply to time-dependent cases. The fisher information of $\lambda^*$ can be decomposed as
\begin{subequations}
\begin{align}
    \operatorname{Var}[\Lambda^*] &= \sum_{i \neq j} \left( \frac{\partial\ln R_{ij}}{\partial \lambda^*} \right)^2 \langle N_{ij} \rangle \\
    &= \sum_\alpha \left( \frac{\partial \lambda^*}{\partial \lambda_\alpha} \right)^2 \sum_{i \neq j} \left( \frac{\partial\ln R_{ij}}{\partial \lambda_\alpha} \right)^2 \langle N_{ij} \rangle \\
    &= \sum_\alpha \left( \frac{\partial \lambda^*}{\partial \lambda_\alpha} \right)^2 \operatorname{Var}[\Lambda_\alpha].
\end{align}
\end{subequations}
Applying the Cauchy-Schwartz inequality $\sum_{\alpha} (x_\alpha / y_\alpha)^2 \ge (\sum_\alpha x_\alpha)^2 / (\sum_\alpha y_\alpha^2)$ yields the following information monotonicity property (choosing $x_\alpha = \frac{\partial \lambda^*}{\partial \lambda_\alpha} \partial_{\lambda_\alpha}\langle Q \rangle$ and $y = \frac{\partial \lambda^*}{\partial \lambda_\alpha} \sqrt{\operatorname{Var}[\Lambda_\alpha]}$):
\begin{equation}
    \frac{(\partial_{\lambda^*} \langle Q \rangle)^2}{\operatorname{Var}[\Lambda^*]} = \frac{\left( \sum_\alpha \frac{\partial \lambda^*}{\partial \lambda_\alpha} \partial_{\lambda_\alpha}\langle Q \rangle \right)^2}{\sum_\alpha \left( \frac{\partial \lambda^*}{\partial \lambda_\alpha} \right)^2 \operatorname{Var}[\Lambda_\alpha]} \le \sum_\alpha \frac{(\partial_{\lambda_{\alpha}} \langle Q \rangle)^2}{\operatorname{Var}[\Lambda_\alpha]}.
\end{equation}

\bibliography{apssamp}

\begin{thebibliography}{57}%
\makeatletter
\providecommand \@ifxundefined [1]{%
 \@ifx{#1\undefined}
}%
\providecommand \@ifnum [1]{%
 \ifnum #1\expandafter \@firstoftwo
 \else \expandafter \@secondoftwo
 \fi
}%
\providecommand \@ifx [1]{%
 \ifx #1\expandafter \@firstoftwo
 \else \expandafter \@secondoftwo
 \fi
}%
\providecommand \natexlab [1]{#1}%
\providecommand \enquote  [1]{``#1''}%
\providecommand \bibnamefont  [1]{#1}%
\providecommand \bibfnamefont [1]{#1}%
\providecommand \citenamefont [1]{#1}%
\providecommand \href@noop [0]{\@secondoftwo}%
\providecommand \href [0]{\begingroup \@sanitize@url \@href}%
\providecommand \@href[1]{\@@startlink{#1}\@@href}%
\providecommand \@@href[1]{\endgroup#1\@@endlink}%
\providecommand \@sanitize@url [0]{\catcode `\\12\catcode `\$12\catcode `\&12\catcode `\#12\catcode `\^12\catcode `\_12\catcode `\%12\relax}%
\providecommand \@@startlink[1]{}%
\providecommand \@@endlink[0]{}%
\providecommand \url  [0]{\begingroup\@sanitize@url \@url }%
\providecommand \@url [1]{\endgroup\@href {#1}{\urlprefix }}%
\providecommand \urlprefix  [0]{URL }%
\providecommand \Eprint [0]{\href }%
\providecommand \doibase [0]{https://doi.org/}%
\providecommand \selectlanguage [0]{\@gobble}%
\providecommand \bibinfo  [0]{\@secondoftwo}%
\providecommand \bibfield  [0]{\@secondoftwo}%
\providecommand \translation [1]{[#1]}%
\providecommand \BibitemOpen [0]{}%
\providecommand \bibitemStop [0]{}%
\providecommand \bibitemNoStop [0]{.\EOS\space}%
\providecommand \EOS [0]{\spacefactor3000\relax}%
\providecommand \BibitemShut  [1]{\csname bibitem#1\endcsname}%
\let\auto@bib@innerbib\@empty
\bibitem [{\citenamefont {Kubo}(1957)}]{kubo1957statistical}%
  \BibitemOpen
  \bibfield  {author} {\bibinfo {author} {\bibfnamefont {R.}~\bibnamefont {Kubo}},\ }\bibfield  {title} {\bibinfo {title} {Statistical-mechanical theory of irreversible processes. i. general theory and simple applications to magnetic and conduction problems},\ }\href@noop {} {\bibfield  {journal} {\bibinfo  {journal} {Journal of the physical society of Japan}\ }\textbf {\bibinfo {volume} {12}},\ \bibinfo {pages} {570} (\bibinfo {year} {1957})}\BibitemShut {NoStop}%
\bibitem [{\citenamefont {Agarwal}(1972)}]{agarwal1972fluctuation}%
  \BibitemOpen
  \bibfield  {author} {\bibinfo {author} {\bibfnamefont {G.~S.}\ \bibnamefont {Agarwal}},\ }\bibfield  {title} {\bibinfo {title} {Fluctuation-dissipation theorems for systems in non-thermal equilibrium and applications},\ }\href@noop {} {\bibfield  {journal} {\bibinfo  {journal} {Zeitschrift f{\"u}r Physik A Hadrons and nuclei}\ }\textbf {\bibinfo {volume} {252}},\ \bibinfo {pages} {25} (\bibinfo {year} {1972})}\BibitemShut {NoStop}%
\bibitem [{\citenamefont {Wu}\ and\ \citenamefont {Wang}(2020)}]{wu2020generalized}%
  \BibitemOpen
  \bibfield  {author} {\bibinfo {author} {\bibfnamefont {W.}~\bibnamefont {Wu}}\ and\ \bibinfo {author} {\bibfnamefont {J.}~\bibnamefont {Wang}},\ }\bibfield  {title} {\bibinfo {title} {Generalized fluctuation-dissipation theorem for non-equilibrium spatially extended systems},\ }\href@noop {} {\bibfield  {journal} {\bibinfo  {journal} {Frontiers in Physics}\ }\textbf {\bibinfo {volume} {8}},\ \bibinfo {pages} {567523} (\bibinfo {year} {2020})}\BibitemShut {NoStop}%
\bibitem [{\citenamefont {Zhang}\ \emph {et~al.}(2021)\citenamefont {Zhang}, \citenamefont {Wang},\ and\ \citenamefont {Wang}}]{zhang2021quantum}%
  \BibitemOpen
  \bibfield  {author} {\bibinfo {author} {\bibfnamefont {Z.}~\bibnamefont {Zhang}}, \bibinfo {author} {\bibfnamefont {X.}~\bibnamefont {Wang}},\ and\ \bibinfo {author} {\bibfnamefont {J.}~\bibnamefont {Wang}},\ }\bibfield  {title} {\bibinfo {title} {Quantum fluctuation-dissipation theorem far from equilibrium},\ }\href@noop {} {\bibfield  {journal} {\bibinfo  {journal} {Physical Review B}\ }\textbf {\bibinfo {volume} {104}},\ \bibinfo {pages} {085439} (\bibinfo {year} {2021})}\BibitemShut {NoStop}%
\bibitem [{\citenamefont {Feng}\ and\ \citenamefont {Wang}(2011)}]{feng2011potential}%
  \BibitemOpen
  \bibfield  {author} {\bibinfo {author} {\bibfnamefont {H.}~\bibnamefont {Feng}}\ and\ \bibinfo {author} {\bibfnamefont {J.}~\bibnamefont {Wang}},\ }\bibfield  {title} {\bibinfo {title} {Potential and flux decomposition for dynamical systems and non-equilibrium thermodynamics: Curvature, gauge field, and generalized fluctuation-dissipation theorem},\ }\href@noop {} {\bibfield  {journal} {\bibinfo  {journal} {The Journal of chemical physics}\ }\textbf {\bibinfo {volume} {135}} (\bibinfo {year} {2011})}\BibitemShut {NoStop}%
\bibitem [{\citenamefont {Marconi}\ \emph {et~al.}(2008)\citenamefont {Marconi}, \citenamefont {Puglisi}, \citenamefont {Rondoni},\ and\ \citenamefont {Vulpiani}}]{marconi2008fluctuation}%
  \BibitemOpen
  \bibfield  {author} {\bibinfo {author} {\bibfnamefont {U.~M.~B.}\ \bibnamefont {Marconi}}, \bibinfo {author} {\bibfnamefont {A.}~\bibnamefont {Puglisi}}, \bibinfo {author} {\bibfnamefont {L.}~\bibnamefont {Rondoni}},\ and\ \bibinfo {author} {\bibfnamefont {A.}~\bibnamefont {Vulpiani}},\ }\bibfield  {title} {\bibinfo {title} {Fluctuation--dissipation: response theory in statistical physics},\ }\href@noop {} {\bibfield  {journal} {\bibinfo  {journal} {Physics reports}\ }\textbf {\bibinfo {volume} {461}},\ \bibinfo {pages} {111} (\bibinfo {year} {2008})}\BibitemShut {NoStop}%
\bibitem [{\citenamefont {Jarzynski}(1997)}]{jarzynski1997nonequilibrium}%
  \BibitemOpen
  \bibfield  {author} {\bibinfo {author} {\bibfnamefont {C.}~\bibnamefont {Jarzynski}},\ }\bibfield  {title} {\bibinfo {title} {Nonequilibrium equality for free energy differences},\ }\href@noop {} {\bibfield  {journal} {\bibinfo  {journal} {Physical Review Letters}\ }\textbf {\bibinfo {volume} {78}},\ \bibinfo {pages} {2690} (\bibinfo {year} {1997})}\BibitemShut {NoStop}%
\bibitem [{\citenamefont {Evans}\ \emph {et~al.}(1993)\citenamefont {Evans}, \citenamefont {Cohen},\ and\ \citenamefont {Morriss}}]{evans1993probability}%
  \BibitemOpen
  \bibfield  {author} {\bibinfo {author} {\bibfnamefont {D.~J.}\ \bibnamefont {Evans}}, \bibinfo {author} {\bibfnamefont {E.~G.~D.}\ \bibnamefont {Cohen}},\ and\ \bibinfo {author} {\bibfnamefont {G.~P.}\ \bibnamefont {Morriss}},\ }\bibfield  {title} {\bibinfo {title} {Probability of second law violations in shearing steady states},\ }\href@noop {} {\bibfield  {journal} {\bibinfo  {journal} {Physical review letters}\ }\textbf {\bibinfo {volume} {71}},\ \bibinfo {pages} {2401} (\bibinfo {year} {1993})}\BibitemShut {NoStop}%
\bibitem [{\citenamefont {Dellago}\ \emph {et~al.}(1998{\natexlab{a}})\citenamefont {Dellago}, \citenamefont {Bolhuis}, \citenamefont {Csajka},\ and\ \citenamefont {Chandler}}]{dellago1998transition}%
  \BibitemOpen
  \bibfield  {author} {\bibinfo {author} {\bibfnamefont {C.}~\bibnamefont {Dellago}}, \bibinfo {author} {\bibfnamefont {P.~G.}\ \bibnamefont {Bolhuis}}, \bibinfo {author} {\bibfnamefont {F.~S.}\ \bibnamefont {Csajka}},\ and\ \bibinfo {author} {\bibfnamefont {D.}~\bibnamefont {Chandler}},\ }\bibfield  {title} {\bibinfo {title} {Transition path sampling and the calculation of rate constants},\ }\href@noop {} {\bibfield  {journal} {\bibinfo  {journal} {The Journal of chemical physics}\ }\textbf {\bibinfo {volume} {108}},\ \bibinfo {pages} {1964} (\bibinfo {year} {1998}{\natexlab{a}})}\BibitemShut {NoStop}%
\bibitem [{\citenamefont {Dellago}\ \emph {et~al.}(1998{\natexlab{b}})\citenamefont {Dellago}, \citenamefont {Bolhuis},\ and\ \citenamefont {Chandler}}]{dellago1998efficient}%
  \BibitemOpen
  \bibfield  {author} {\bibinfo {author} {\bibfnamefont {C.}~\bibnamefont {Dellago}}, \bibinfo {author} {\bibfnamefont {P.~G.}\ \bibnamefont {Bolhuis}},\ and\ \bibinfo {author} {\bibfnamefont {D.}~\bibnamefont {Chandler}},\ }\bibfield  {title} {\bibinfo {title} {Efficient transition path sampling: Application to lennard-jones cluster rearrangements},\ }\href@noop {} {\bibfield  {journal} {\bibinfo  {journal} {The Journal of chemical physics}\ }\textbf {\bibinfo {volume} {108}},\ \bibinfo {pages} {9236} (\bibinfo {year} {1998}{\natexlab{b}})}\BibitemShut {NoStop}%
\bibitem [{\citenamefont {Bolhuis}\ \emph {et~al.}(2002)\citenamefont {Bolhuis}, \citenamefont {Chandler}, \citenamefont {Dellago},\ and\ \citenamefont {Geissler}}]{bolhuis2002transition}%
  \BibitemOpen
  \bibfield  {author} {\bibinfo {author} {\bibfnamefont {P.~G.}\ \bibnamefont {Bolhuis}}, \bibinfo {author} {\bibfnamefont {D.}~\bibnamefont {Chandler}}, \bibinfo {author} {\bibfnamefont {C.}~\bibnamefont {Dellago}},\ and\ \bibinfo {author} {\bibfnamefont {P.~L.}\ \bibnamefont {Geissler}},\ }\bibfield  {title} {\bibinfo {title} {Transition path sampling: Throwing ropes over rough mountain passes, in the dark},\ }\href@noop {} {\bibfield  {journal} {\bibinfo  {journal} {Annual review of physical chemistry}\ }\textbf {\bibinfo {volume} {53}},\ \bibinfo {pages} {291} (\bibinfo {year} {2002})}\BibitemShut {NoStop}%
\bibitem [{\citenamefont {Touchette}(2009)}]{touchette2009large}%
  \BibitemOpen
  \bibfield  {author} {\bibinfo {author} {\bibfnamefont {H.}~\bibnamefont {Touchette}},\ }\bibfield  {title} {\bibinfo {title} {The large deviation approach to statistical mechanics},\ }\href@noop {} {\bibfield  {journal} {\bibinfo  {journal} {Physics Reports}\ }\textbf {\bibinfo {volume} {478}},\ \bibinfo {pages} {1} (\bibinfo {year} {2009})}\BibitemShut {NoStop}%
\bibitem [{\citenamefont {Chetrite}\ and\ \citenamefont {Touchette}(2013)}]{chetrite2013nonequilibrium}%
  \BibitemOpen
  \bibfield  {author} {\bibinfo {author} {\bibfnamefont {R.}~\bibnamefont {Chetrite}}\ and\ \bibinfo {author} {\bibfnamefont {H.}~\bibnamefont {Touchette}},\ }\bibfield  {title} {\bibinfo {title} {Nonequilibrium microcanonical and canonical ensembles and their equivalence},\ }\href@noop {} {\bibfield  {journal} {\bibinfo  {journal} {Physical review letters}\ }\textbf {\bibinfo {volume} {111}},\ \bibinfo {pages} {120601} (\bibinfo {year} {2013})}\BibitemShut {NoStop}%
\bibitem [{\citenamefont {Seifert}(2012)}]{seifert2012stochastic}%
  \BibitemOpen
  \bibfield  {author} {\bibinfo {author} {\bibfnamefont {U.}~\bibnamefont {Seifert}},\ }\bibfield  {title} {\bibinfo {title} {Stochastic thermodynamics, fluctuation theorems and molecular machines},\ }\href@noop {} {\bibfield  {journal} {\bibinfo  {journal} {Reports on progress in physics}\ }\textbf {\bibinfo {volume} {75}},\ \bibinfo {pages} {126001} (\bibinfo {year} {2012})}\BibitemShut {NoStop}%
\bibitem [{\citenamefont {Bertini}\ \emph {et~al.}(2015)\citenamefont {Bertini}, \citenamefont {De~Sole}, \citenamefont {Gabrielli}, \citenamefont {Jona-Lasinio},\ and\ \citenamefont {Landim}}]{bertini2015macroscopic}%
  \BibitemOpen
  \bibfield  {author} {\bibinfo {author} {\bibfnamefont {L.}~\bibnamefont {Bertini}}, \bibinfo {author} {\bibfnamefont {A.}~\bibnamefont {De~Sole}}, \bibinfo {author} {\bibfnamefont {D.}~\bibnamefont {Gabrielli}}, \bibinfo {author} {\bibfnamefont {G.}~\bibnamefont {Jona-Lasinio}},\ and\ \bibinfo {author} {\bibfnamefont {C.}~\bibnamefont {Landim}},\ }\bibfield  {title} {\bibinfo {title} {Macroscopic fluctuation theory},\ }\href@noop {} {\bibfield  {journal} {\bibinfo  {journal} {Reviews of Modern Physics}\ }\textbf {\bibinfo {volume} {87}},\ \bibinfo {pages} {593} (\bibinfo {year} {2015})}\BibitemShut {NoStop}%
\bibitem [{\citenamefont {Maes}(2020{\natexlab{a}})}]{maes2020frenesy}%
  \BibitemOpen
  \bibfield  {author} {\bibinfo {author} {\bibfnamefont {C.}~\bibnamefont {Maes}},\ }\bibfield  {title} {\bibinfo {title} {Frenesy: Time-symmetric dynamical activity in nonequilibria},\ }\href@noop {} {\bibfield  {journal} {\bibinfo  {journal} {Physics Reports}\ }\textbf {\bibinfo {volume} {850}},\ \bibinfo {pages} {1} (\bibinfo {year} {2020}{\natexlab{a}})}\BibitemShut {NoStop}%
\bibitem [{\citenamefont {Seifert}\ and\ \citenamefont {Speck}(2010)}]{seifert2010fluctuation}%
  \BibitemOpen
  \bibfield  {author} {\bibinfo {author} {\bibfnamefont {U.}~\bibnamefont {Seifert}}\ and\ \bibinfo {author} {\bibfnamefont {T.}~\bibnamefont {Speck}},\ }\bibfield  {title} {\bibinfo {title} {Fluctuation-dissipation theorem in nonequilibrium steady states},\ }\href@noop {} {\bibfield  {journal} {\bibinfo  {journal} {Europhysics Letters}\ }\textbf {\bibinfo {volume} {89}},\ \bibinfo {pages} {10007} (\bibinfo {year} {2010})}\BibitemShut {NoStop}%
\bibitem [{\citenamefont {Baiesi}\ \emph {et~al.}(2009)\citenamefont {Baiesi}, \citenamefont {Maes},\ and\ \citenamefont {Wynants}}]{baiesi2009nonequilibrium}%
  \BibitemOpen
  \bibfield  {author} {\bibinfo {author} {\bibfnamefont {M.}~\bibnamefont {Baiesi}}, \bibinfo {author} {\bibfnamefont {C.}~\bibnamefont {Maes}},\ and\ \bibinfo {author} {\bibfnamefont {B.}~\bibnamefont {Wynants}},\ }\bibfield  {title} {\bibinfo {title} {Nonequilibrium linear response for markov dynamics, i: jump processes and overdamped diffusions},\ }\href@noop {} {\bibfield  {journal} {\bibinfo  {journal} {Journal of statistical physics}\ }\textbf {\bibinfo {volume} {137}},\ \bibinfo {pages} {1094} (\bibinfo {year} {2009})}\BibitemShut {NoStop}%
\bibitem [{\citenamefont {Baiesi}\ and\ \citenamefont {Maes}(2013)}]{baiesi2013update}%
  \BibitemOpen
  \bibfield  {author} {\bibinfo {author} {\bibfnamefont {M.}~\bibnamefont {Baiesi}}\ and\ \bibinfo {author} {\bibfnamefont {C.}~\bibnamefont {Maes}},\ }\bibfield  {title} {\bibinfo {title} {An update on the nonequilibrium linear response},\ }\href@noop {} {\bibfield  {journal} {\bibinfo  {journal} {New Journal of Physics}\ }\textbf {\bibinfo {volume} {15}},\ \bibinfo {pages} {013004} (\bibinfo {year} {2013})}\BibitemShut {NoStop}%
\bibitem [{\citenamefont {Maes}(2020{\natexlab{b}})}]{maes2020response}%
  \BibitemOpen
  \bibfield  {author} {\bibinfo {author} {\bibfnamefont {C.}~\bibnamefont {Maes}},\ }\bibfield  {title} {\bibinfo {title} {Response theory: a trajectory-based approach},\ }\href@noop {} {\bibfield  {journal} {\bibinfo  {journal} {Frontiers in Physics}\ }\textbf {\bibinfo {volume} {8}},\ \bibinfo {pages} {229} (\bibinfo {year} {2020}{\natexlab{b}})}\BibitemShut {NoStop}%
\bibitem [{\citenamefont {Dechant}\ and\ \citenamefont {Sasa}(2020)}]{dechant2020fluctuation}%
  \BibitemOpen
  \bibfield  {author} {\bibinfo {author} {\bibfnamefont {A.}~\bibnamefont {Dechant}}\ and\ \bibinfo {author} {\bibfnamefont {S.-i.}\ \bibnamefont {Sasa}},\ }\bibfield  {title} {\bibinfo {title} {Fluctuation--response inequality out of equilibrium},\ }\href@noop {} {\bibfield  {journal} {\bibinfo  {journal} {Proceedings of the National Academy of Sciences}\ }\textbf {\bibinfo {volume} {117}},\ \bibinfo {pages} {6430} (\bibinfo {year} {2020})}\BibitemShut {NoStop}%
\bibitem [{\citenamefont {Zheng}\ and\ \citenamefont {Lu}(2024)}]{zheng2024universal}%
  \BibitemOpen
  \bibfield  {author} {\bibinfo {author} {\bibfnamefont {J.}~\bibnamefont {Zheng}}\ and\ \bibinfo {author} {\bibfnamefont {Z.}~\bibnamefont {Lu}},\ }\bibfield  {title} {\bibinfo {title} {Universal non-equilibrium response theory beyond steady states},\ }\href@noop {} {\bibfield  {journal} {\bibinfo  {journal} {arXiv preprint arXiv:2403.10952}\ } (\bibinfo {year} {2024})}\BibitemShut {NoStop}%
\bibitem [{\citenamefont {Kwon}\ \emph {et~al.}(2024)\citenamefont {Kwon}, \citenamefont {Chun}, \citenamefont {Park},\ and\ \citenamefont {Lee}}]{kwon2024fluctuation}%
  \BibitemOpen
  \bibfield  {author} {\bibinfo {author} {\bibfnamefont {E.}~\bibnamefont {Kwon}}, \bibinfo {author} {\bibfnamefont {H.-M.}\ \bibnamefont {Chun}}, \bibinfo {author} {\bibfnamefont {H.}~\bibnamefont {Park}},\ and\ \bibinfo {author} {\bibfnamefont {J.~S.}\ \bibnamefont {Lee}},\ }\bibfield  {title} {\bibinfo {title} {Fluctuation-response inequalities for kinetic and entropic perturbations},\ }\href@noop {} {\bibfield  {journal} {\bibinfo  {journal} {arXiv preprint arXiv:2411.18108}\ } (\bibinfo {year} {2024})}\BibitemShut {NoStop}%
\bibitem [{\citenamefont {Ptaszy{\'n}ski}\ \emph {et~al.}(2024)\citenamefont {Ptaszy{\'n}ski}, \citenamefont {Aslyamov},\ and\ \citenamefont {Esposito}}]{ptaszynski2024dissipation}%
  \BibitemOpen
  \bibfield  {author} {\bibinfo {author} {\bibfnamefont {K.}~\bibnamefont {Ptaszy{\'n}ski}}, \bibinfo {author} {\bibfnamefont {T.}~\bibnamefont {Aslyamov}},\ and\ \bibinfo {author} {\bibfnamefont {M.}~\bibnamefont {Esposito}},\ }\bibfield  {title} {\bibinfo {title} {Dissipation bounds precision of current response to kinetic perturbations},\ }\href@noop {} {\bibfield  {journal} {\bibinfo  {journal} {Physical Review Letters}\ }\textbf {\bibinfo {volume} {133}},\ \bibinfo {pages} {227101} (\bibinfo {year} {2024})}\BibitemShut {NoStop}%
\bibitem [{\citenamefont {Aslyamov}\ \emph {et~al.}(2025)\citenamefont {Aslyamov}, \citenamefont {Ptaszy{\'n}ski},\ and\ \citenamefont {Esposito}}]{aslyamov2025nonequilibrium}%
  \BibitemOpen
  \bibfield  {author} {\bibinfo {author} {\bibfnamefont {T.}~\bibnamefont {Aslyamov}}, \bibinfo {author} {\bibfnamefont {K.}~\bibnamefont {Ptaszy{\'n}ski}},\ and\ \bibinfo {author} {\bibfnamefont {M.}~\bibnamefont {Esposito}},\ }\bibfield  {title} {\bibinfo {title} {Nonequilibrium fluctuation-response relations: From identities to bounds},\ }\href@noop {} {\bibfield  {journal} {\bibinfo  {journal} {Physical Review Letters}\ }\textbf {\bibinfo {volume} {134}},\ \bibinfo {pages} {157101} (\bibinfo {year} {2025})}\BibitemShut {NoStop}%
\bibitem [{\citenamefont {Ptaszynski}\ \emph {et~al.}(2024)\citenamefont {Ptaszynski}, \citenamefont {Aslyamov},\ and\ \citenamefont {Esposito}}]{ptaszynski2024nonequilibrium}%
  \BibitemOpen
  \bibfield  {author} {\bibinfo {author} {\bibfnamefont {K.}~\bibnamefont {Ptaszynski}}, \bibinfo {author} {\bibfnamefont {T.}~\bibnamefont {Aslyamov}},\ and\ \bibinfo {author} {\bibfnamefont {M.}~\bibnamefont {Esposito}},\ }\bibfield  {title} {\bibinfo {title} {Nonequilibrium fluctuation-response relations for state observables},\ }\href@noop {} {\bibfield  {journal} {\bibinfo  {journal} {arXiv preprint arXiv:2412.10233}\ } (\bibinfo {year} {2024})}\BibitemShut {NoStop}%
\bibitem [{\citenamefont {Ptaszynski}\ \emph {et~al.}(2025)\citenamefont {Ptaszynski}, \citenamefont {Aslyamov},\ and\ \citenamefont {Esposito}}]{ptaszynski2025nonequilibrium}%
  \BibitemOpen
  \bibfield  {author} {\bibinfo {author} {\bibfnamefont {K.}~\bibnamefont {Ptaszynski}}, \bibinfo {author} {\bibfnamefont {T.}~\bibnamefont {Aslyamov}},\ and\ \bibinfo {author} {\bibfnamefont {M.}~\bibnamefont {Esposito}},\ }\bibfield  {title} {\bibinfo {title} {Nonequilibrium fluctuation-response relations for state-current correlations},\ }\href@noop {} {\bibfield  {journal} {\bibinfo  {journal} {arXiv preprint arXiv:2506.08877}\ } (\bibinfo {year} {2025})}\BibitemShut {NoStop}%
\bibitem [{\citenamefont {Liu}\ and\ \citenamefont {Gu}(2025)}]{liu2025dynamical}%
  \BibitemOpen
  \bibfield  {author} {\bibinfo {author} {\bibfnamefont {K.}~\bibnamefont {Liu}}\ and\ \bibinfo {author} {\bibfnamefont {J.}~\bibnamefont {Gu}},\ }\bibfield  {title} {\bibinfo {title} {Dynamical activity universally bounds precision of response in markovian nonequilibrium systems},\ }\href@noop {} {\bibfield  {journal} {\bibinfo  {journal} {Communications Physics}\ }\textbf {\bibinfo {volume} {8}},\ \bibinfo {pages} {62} (\bibinfo {year} {2025})}\BibitemShut {NoStop}%
\bibitem [{\citenamefont {Harunari}\ \emph {et~al.}(2024)\citenamefont {Harunari}, \citenamefont {Dal~Cengio}, \citenamefont {Lecomte},\ and\ \citenamefont {Polettini}}]{harunari2024mutual}%
  \BibitemOpen
  \bibfield  {author} {\bibinfo {author} {\bibfnamefont {P.~E.}\ \bibnamefont {Harunari}}, \bibinfo {author} {\bibfnamefont {S.}~\bibnamefont {Dal~Cengio}}, \bibinfo {author} {\bibfnamefont {V.}~\bibnamefont {Lecomte}},\ and\ \bibinfo {author} {\bibfnamefont {M.}~\bibnamefont {Polettini}},\ }\bibfield  {title} {\bibinfo {title} {Mutual linearity of nonequilibrium network currents},\ }\href@noop {} {\bibfield  {journal} {\bibinfo  {journal} {Physical Review Letters}\ }\textbf {\bibinfo {volume} {133}},\ \bibinfo {pages} {047401} (\bibinfo {year} {2024})}\BibitemShut {NoStop}%
\bibitem [{\citenamefont {Owen}\ \emph {et~al.}(2020)\citenamefont {Owen}, \citenamefont {Gingrich},\ and\ \citenamefont {Horowitz}}]{owen2020universal}%
  \BibitemOpen
  \bibfield  {author} {\bibinfo {author} {\bibfnamefont {J.~A.}\ \bibnamefont {Owen}}, \bibinfo {author} {\bibfnamefont {T.~R.}\ \bibnamefont {Gingrich}},\ and\ \bibinfo {author} {\bibfnamefont {J.~M.}\ \bibnamefont {Horowitz}},\ }\bibfield  {title} {\bibinfo {title} {Universal thermodynamic bounds on nonequilibrium response with biochemical applications},\ }\href@noop {} {\bibfield  {journal} {\bibinfo  {journal} {Physical Review X}\ }\textbf {\bibinfo {volume} {10}},\ \bibinfo {pages} {011066} (\bibinfo {year} {2020})}\BibitemShut {NoStop}%
\bibitem [{\citenamefont {Owen}\ and\ \citenamefont {Horowitz}(2023)}]{owen2023size}%
  \BibitemOpen
  \bibfield  {author} {\bibinfo {author} {\bibfnamefont {J.~A.}\ \bibnamefont {Owen}}\ and\ \bibinfo {author} {\bibfnamefont {J.~M.}\ \bibnamefont {Horowitz}},\ }\bibfield  {title} {\bibinfo {title} {Size limits the sensitivity of kinetic schemes},\ }\href@noop {} {\bibfield  {journal} {\bibinfo  {journal} {Nature communications}\ }\textbf {\bibinfo {volume} {14}},\ \bibinfo {pages} {1280} (\bibinfo {year} {2023})}\BibitemShut {NoStop}%
\bibitem [{\citenamefont {L{\'e}vy}(1940)}]{levy1940certains}%
  \BibitemOpen
  \bibfield  {author} {\bibinfo {author} {\bibfnamefont {P.}~\bibnamefont {L{\'e}vy}},\ }\bibfield  {title} {\bibinfo {title} {Sur certains processus stochastiques homog{\`e}nes},\ }\href@noop {} {\bibfield  {journal} {\bibinfo  {journal} {Compositio mathematica}\ }\textbf {\bibinfo {volume} {7}},\ \bibinfo {pages} {283} (\bibinfo {year} {1940})}\BibitemShut {NoStop}%
\bibitem [{\citenamefont {Godreche}\ and\ \citenamefont {Luck}(2001)}]{godreche2001statistics}%
  \BibitemOpen
  \bibfield  {author} {\bibinfo {author} {\bibfnamefont {C.}~\bibnamefont {Godreche}}\ and\ \bibinfo {author} {\bibfnamefont {J.}~\bibnamefont {Luck}},\ }\bibfield  {title} {\bibinfo {title} {Statistics of the occupation time of renewal processes},\ }\href@noop {} {\bibfield  {journal} {\bibinfo  {journal} {Journal of Statistical Physics}\ }\textbf {\bibinfo {volume} {104}},\ \bibinfo {pages} {489} (\bibinfo {year} {2001})}\BibitemShut {NoStop}%
\bibitem [{\citenamefont {Majumdar}\ and\ \citenamefont {Comtet}(2002)}]{majumdar2002local}%
  \BibitemOpen
  \bibfield  {author} {\bibinfo {author} {\bibfnamefont {S.~N.}\ \bibnamefont {Majumdar}}\ and\ \bibinfo {author} {\bibfnamefont {A.}~\bibnamefont {Comtet}},\ }\bibfield  {title} {\bibinfo {title} {Local and occupation time of a particle diffusing in a random medium},\ }\href@noop {} {\bibfield  {journal} {\bibinfo  {journal} {Physical review letters}\ }\textbf {\bibinfo {volume} {89}},\ \bibinfo {pages} {060601} (\bibinfo {year} {2002})}\BibitemShut {NoStop}%
\bibitem [{\citenamefont {Lapolla}\ and\ \citenamefont {Godec}(2019)}]{lapolla2019manifestations}%
  \BibitemOpen
  \bibfield  {author} {\bibinfo {author} {\bibfnamefont {A.}~\bibnamefont {Lapolla}}\ and\ \bibinfo {author} {\bibfnamefont {A.}~\bibnamefont {Godec}},\ }\bibfield  {title} {\bibinfo {title} {Manifestations of projection-induced memory: General theory and the tilted single file},\ }\href@noop {} {\bibfield  {journal} {\bibinfo  {journal} {Frontiers in Physics}\ }\textbf {\bibinfo {volume} {7}},\ \bibinfo {pages} {182} (\bibinfo {year} {2019})}\BibitemShut {NoStop}%
\bibitem [{\citenamefont {Hartich}\ and\ \citenamefont {Godec}(2023)}]{hartich2023violation}%
  \BibitemOpen
  \bibfield  {author} {\bibinfo {author} {\bibfnamefont {D.}~\bibnamefont {Hartich}}\ and\ \bibinfo {author} {\bibfnamefont {A.}~\bibnamefont {Godec}},\ }\bibfield  {title} {\bibinfo {title} {Violation of local detailed balance upon lumping despite a clear timescale separation},\ }\href@noop {} {\bibfield  {journal} {\bibinfo  {journal} {Physical Review Research}\ }\textbf {\bibinfo {volume} {5}},\ \bibinfo {pages} {L032017} (\bibinfo {year} {2023})}\BibitemShut {NoStop}%
\bibitem [{\citenamefont {Ramaswamy}(2010)}]{ramaswamy2010mechanics}%
  \BibitemOpen
  \bibfield  {author} {\bibinfo {author} {\bibfnamefont {S.}~\bibnamefont {Ramaswamy}},\ }\bibfield  {title} {\bibinfo {title} {The mechanics and statistics of active matter},\ }\href@noop {} {\bibfield  {journal} {\bibinfo  {journal} {Annu. Rev. Condens. Matter Phys.}\ }\textbf {\bibinfo {volume} {1}},\ \bibinfo {pages} {323} (\bibinfo {year} {2010})}\BibitemShut {NoStop}%
\bibitem [{\citenamefont {Ramaswamy}(2017)}]{ramaswamy2017active}%
  \BibitemOpen
  \bibfield  {author} {\bibinfo {author} {\bibfnamefont {S.}~\bibnamefont {Ramaswamy}},\ }\bibfield  {title} {\bibinfo {title} {Active matter},\ }\href@noop {} {\bibfield  {journal} {\bibinfo  {journal} {Journal of Statistical Mechanics: Theory and Experiment}\ }\textbf {\bibinfo {volume} {2017}},\ \bibinfo {pages} {054002} (\bibinfo {year} {2017})}\BibitemShut {NoStop}%
\bibitem [{\citenamefont {Marchetti}\ \emph {et~al.}(2013)\citenamefont {Marchetti}, \citenamefont {Joanny}, \citenamefont {Ramaswamy}, \citenamefont {Liverpool}, \citenamefont {Prost}, \citenamefont {Rao},\ and\ \citenamefont {Simha}}]{marchetti2013hydrodynamics}%
  \BibitemOpen
  \bibfield  {author} {\bibinfo {author} {\bibfnamefont {M.~C.}\ \bibnamefont {Marchetti}}, \bibinfo {author} {\bibfnamefont {J.-F.}\ \bibnamefont {Joanny}}, \bibinfo {author} {\bibfnamefont {S.}~\bibnamefont {Ramaswamy}}, \bibinfo {author} {\bibfnamefont {T.~B.}\ \bibnamefont {Liverpool}}, \bibinfo {author} {\bibfnamefont {J.}~\bibnamefont {Prost}}, \bibinfo {author} {\bibfnamefont {M.}~\bibnamefont {Rao}},\ and\ \bibinfo {author} {\bibfnamefont {R.~A.}\ \bibnamefont {Simha}},\ }\bibfield  {title} {\bibinfo {title} {Hydrodynamics of soft active matter},\ }\href@noop {} {\bibfield  {journal} {\bibinfo  {journal} {Reviews of modern physics}\ }\textbf {\bibinfo {volume} {85}},\ \bibinfo {pages} {1143} (\bibinfo {year} {2013})}\BibitemShut {NoStop}%
\bibitem [{\citenamefont {O'Connor}(2012)}]{o2012time}%
  \BibitemOpen
  \bibfield  {author} {\bibinfo {author} {\bibfnamefont {D.}~\bibnamefont {O'Connor}},\ }\href@noop {} {\emph {\bibinfo {title} {Time-correlated single photon counting}}}\ (\bibinfo  {publisher} {Academic press},\ \bibinfo {year} {2012})\BibitemShut {NoStop}%
\bibitem [{\citenamefont {Margolin}\ and\ \citenamefont {Barkai}(2005)}]{margolin2005nonergodicity}%
  \BibitemOpen
  \bibfield  {author} {\bibinfo {author} {\bibfnamefont {G.}~\bibnamefont {Margolin}}\ and\ \bibinfo {author} {\bibfnamefont {E.}~\bibnamefont {Barkai}},\ }\bibfield  {title} {\bibinfo {title} {Nonergodicity of blinking nanocrystals and other l{\'e}vy-walk processes},\ }\href@noop {} {\bibfield  {journal} {\bibinfo  {journal} {Physical review letters}\ }\textbf {\bibinfo {volume} {94}},\ \bibinfo {pages} {080601} (\bibinfo {year} {2005})}\BibitemShut {NoStop}%
\bibitem [{\citenamefont {Ramesh}\ \emph {et~al.}(2024)\citenamefont {Ramesh}, \citenamefont {Peters},\ and\ \citenamefont {Rodriguez}}]{ramesh2024arcsine}%
  \BibitemOpen
  \bibfield  {author} {\bibinfo {author} {\bibfnamefont {V.}~\bibnamefont {Ramesh}}, \bibinfo {author} {\bibfnamefont {K.}~\bibnamefont {Peters}},\ and\ \bibinfo {author} {\bibfnamefont {S.}~\bibnamefont {Rodriguez}},\ }\bibfield  {title} {\bibinfo {title} {Arcsine laws of light},\ }\href@noop {} {\bibfield  {journal} {\bibinfo  {journal} {Physical Review Letters}\ }\textbf {\bibinfo {volume} {132}},\ \bibinfo {pages} {133801} (\bibinfo {year} {2024})}\BibitemShut {NoStop}%
\bibitem [{\citenamefont {Gopich}\ and\ \citenamefont {Szabo}(2012)}]{gopich2012theory}%
  \BibitemOpen
  \bibfield  {author} {\bibinfo {author} {\bibfnamefont {I.~V.}\ \bibnamefont {Gopich}}\ and\ \bibinfo {author} {\bibfnamefont {A.}~\bibnamefont {Szabo}},\ }\bibfield  {title} {\bibinfo {title} {Theory of the energy transfer efficiency and fluorescence lifetime distribution in single-molecule fret},\ }\href@noop {} {\bibfield  {journal} {\bibinfo  {journal} {Proceedings of the National Academy of Sciences}\ }\textbf {\bibinfo {volume} {109}},\ \bibinfo {pages} {7747} (\bibinfo {year} {2012})}\BibitemShut {NoStop}%
\bibitem [{\citenamefont {Bialek}\ and\ \citenamefont {Setayeshgar}(2005)}]{bialek2005physical}%
  \BibitemOpen
  \bibfield  {author} {\bibinfo {author} {\bibfnamefont {W.}~\bibnamefont {Bialek}}\ and\ \bibinfo {author} {\bibfnamefont {S.}~\bibnamefont {Setayeshgar}},\ }\bibfield  {title} {\bibinfo {title} {Physical limits to biochemical signaling},\ }\href@noop {} {\bibfield  {journal} {\bibinfo  {journal} {Proceedings of the National Academy of Sciences}\ }\textbf {\bibinfo {volume} {102}},\ \bibinfo {pages} {10040} (\bibinfo {year} {2005})}\BibitemShut {NoStop}%
\bibitem [{\citenamefont {Endres}\ and\ \citenamefont {Wingreen}(2008)}]{endres2008accuracy}%
  \BibitemOpen
  \bibfield  {author} {\bibinfo {author} {\bibfnamefont {R.~G.}\ \bibnamefont {Endres}}\ and\ \bibinfo {author} {\bibfnamefont {N.~S.}\ \bibnamefont {Wingreen}},\ }\bibfield  {title} {\bibinfo {title} {Accuracy of direct gradient sensing by single cells},\ }\href@noop {} {\bibfield  {journal} {\bibinfo  {journal} {Proceedings of the National Academy of Sciences}\ }\textbf {\bibinfo {volume} {105}},\ \bibinfo {pages} {15749} (\bibinfo {year} {2008})}\BibitemShut {NoStop}%
\bibitem [{\citenamefont {Mora}\ and\ \citenamefont {Wingreen}(2010)}]{mora2010limits}%
  \BibitemOpen
  \bibfield  {author} {\bibinfo {author} {\bibfnamefont {T.}~\bibnamefont {Mora}}\ and\ \bibinfo {author} {\bibfnamefont {N.~S.}\ \bibnamefont {Wingreen}},\ }\bibfield  {title} {\bibinfo {title} {Limits of sensing temporal concentration changes by single cells},\ }\href@noop {} {\bibfield  {journal} {\bibinfo  {journal} {Physical review letters}\ }\textbf {\bibinfo {volume} {104}},\ \bibinfo {pages} {248101} (\bibinfo {year} {2010})}\BibitemShut {NoStop}%
\bibitem [{\citenamefont {Govern}\ and\ \citenamefont {ten Wolde}(2012)}]{govern2012fundamental}%
  \BibitemOpen
  \bibfield  {author} {\bibinfo {author} {\bibfnamefont {C.~C.}\ \bibnamefont {Govern}}\ and\ \bibinfo {author} {\bibfnamefont {P.~R.}\ \bibnamefont {ten Wolde}},\ }\bibfield  {title} {\bibinfo {title} {Fundamental limits on sensing chemical concentrations with linear biochemical networks},\ }\href@noop {} {\bibfield  {journal} {\bibinfo  {journal} {Physical review letters}\ }\textbf {\bibinfo {volume} {109}},\ \bibinfo {pages} {218103} (\bibinfo {year} {2012})}\BibitemShut {NoStop}%
\bibitem [{\citenamefont {Govern}\ and\ \citenamefont {ten Wolde}(2014)}]{govern2014energy}%
  \BibitemOpen
  \bibfield  {author} {\bibinfo {author} {\bibfnamefont {C.~C.}\ \bibnamefont {Govern}}\ and\ \bibinfo {author} {\bibfnamefont {P.~R.}\ \bibnamefont {ten Wolde}},\ }\bibfield  {title} {\bibinfo {title} {Energy dissipation and noise correlations in biochemical sensing},\ }\href@noop {} {\bibfield  {journal} {\bibinfo  {journal} {Physical review letters}\ }\textbf {\bibinfo {volume} {113}},\ \bibinfo {pages} {258102} (\bibinfo {year} {2014})}\BibitemShut {NoStop}%
\bibitem [{\citenamefont {Harvey}\ \emph {et~al.}(2023)\citenamefont {Harvey}, \citenamefont {Lahiri},\ and\ \citenamefont {Ganguli}}]{harvey2023universal}%
  \BibitemOpen
  \bibfield  {author} {\bibinfo {author} {\bibfnamefont {S.~E.}\ \bibnamefont {Harvey}}, \bibinfo {author} {\bibfnamefont {S.}~\bibnamefont {Lahiri}},\ and\ \bibinfo {author} {\bibfnamefont {S.}~\bibnamefont {Ganguli}},\ }\bibfield  {title} {\bibinfo {title} {Universal energy-accuracy tradeoffs in nonequilibrium cellular sensing},\ }\href@noop {} {\bibfield  {journal} {\bibinfo  {journal} {Physical Review E}\ }\textbf {\bibinfo {volume} {108}},\ \bibinfo {pages} {014403} (\bibinfo {year} {2023})}\BibitemShut {NoStop}%
\bibitem [{\citenamefont {Maffeo}\ \emph {et~al.}(2012)\citenamefont {Maffeo}, \citenamefont {Bhattacharya}, \citenamefont {Yoo}, \citenamefont {Wells},\ and\ \citenamefont {Aksimentiev}}]{maffeo2012modeling}%
  \BibitemOpen
  \bibfield  {author} {\bibinfo {author} {\bibfnamefont {C.}~\bibnamefont {Maffeo}}, \bibinfo {author} {\bibfnamefont {S.}~\bibnamefont {Bhattacharya}}, \bibinfo {author} {\bibfnamefont {J.}~\bibnamefont {Yoo}}, \bibinfo {author} {\bibfnamefont {D.}~\bibnamefont {Wells}},\ and\ \bibinfo {author} {\bibfnamefont {A.}~\bibnamefont {Aksimentiev}},\ }\bibfield  {title} {\bibinfo {title} {Modeling and simulation of ion channels},\ }\href@noop {} {\bibfield  {journal} {\bibinfo  {journal} {Chemical reviews}\ }\textbf {\bibinfo {volume} {112}},\ \bibinfo {pages} {6250} (\bibinfo {year} {2012})}\BibitemShut {NoStop}%
\bibitem [{\citenamefont {Catterall}(2010)}]{catterall2010ion}%
  \BibitemOpen
  \bibfield  {author} {\bibinfo {author} {\bibfnamefont {W.~A.}\ \bibnamefont {Catterall}},\ }\bibfield  {title} {\bibinfo {title} {Ion channel voltage sensors: structure, function, and pathophysiology},\ }\href@noop {} {\bibfield  {journal} {\bibinfo  {journal} {Neuron}\ }\textbf {\bibinfo {volume} {67}},\ \bibinfo {pages} {915} (\bibinfo {year} {2010})}\BibitemShut {NoStop}%
\bibitem [{\citenamefont {Roux}\ \emph {et~al.}(2004)\citenamefont {Roux}, \citenamefont {Allen}, \citenamefont {Berneche},\ and\ \citenamefont {Im}}]{roux2004theoretical}%
  \BibitemOpen
  \bibfield  {author} {\bibinfo {author} {\bibfnamefont {B.}~\bibnamefont {Roux}}, \bibinfo {author} {\bibfnamefont {T.}~\bibnamefont {Allen}}, \bibinfo {author} {\bibfnamefont {S.}~\bibnamefont {Berneche}},\ and\ \bibinfo {author} {\bibfnamefont {W.}~\bibnamefont {Im}},\ }\bibfield  {title} {\bibinfo {title} {Theoretical and computational models of biological ion channels},\ }\href@noop {} {\bibfield  {journal} {\bibinfo  {journal} {Quarterly reviews of biophysics}\ }\textbf {\bibinfo {volume} {37}},\ \bibinfo {pages} {15} (\bibinfo {year} {2004})}\BibitemShut {NoStop}%
\bibitem [{\citenamefont {Peliti}\ and\ \citenamefont {Pigolotti}(2021)}]{peliti2021stochastic}%
  \BibitemOpen
  \bibfield  {author} {\bibinfo {author} {\bibfnamefont {L.}~\bibnamefont {Peliti}}\ and\ \bibinfo {author} {\bibfnamefont {S.}~\bibnamefont {Pigolotti}},\ }\href@noop {} {\emph {\bibinfo {title} {Stochastic thermodynamics: an introduction}}}\ (\bibinfo  {publisher} {Princeton University Press},\ \bibinfo {year} {2021})\BibitemShut {NoStop}%
\bibitem [{\citenamefont {Aslyamov}\ and\ \citenamefont {Esposito}(2024)}]{aslyamov2024nonequilibrium}%
  \BibitemOpen
  \bibfield  {author} {\bibinfo {author} {\bibfnamefont {T.}~\bibnamefont {Aslyamov}}\ and\ \bibinfo {author} {\bibfnamefont {M.}~\bibnamefont {Esposito}},\ }\bibfield  {title} {\bibinfo {title} {Nonequilibrium response for markov jump processes: exact results and tight bounds},\ }\href@noop {} {\bibfield  {journal} {\bibinfo  {journal} {Physical Review Letters}\ }\textbf {\bibinfo {volume} {132}},\ \bibinfo {pages} {037101} (\bibinfo {year} {2024})}\BibitemShut {NoStop}%
\bibitem [{\citenamefont {Dechant}(2018)}]{dechant2018multidimensional}%
  \BibitemOpen
  \bibfield  {author} {\bibinfo {author} {\bibfnamefont {A.}~\bibnamefont {Dechant}},\ }\bibfield  {title} {\bibinfo {title} {Multidimensional thermodynamic uncertainty relations},\ }\href@noop {} {\bibfield  {journal} {\bibinfo  {journal} {Journal of Physics A: Mathematical and Theoretical}\ }\textbf {\bibinfo {volume} {52}},\ \bibinfo {pages} {035001} (\bibinfo {year} {2018})}\BibitemShut {NoStop}%
\bibitem [{\citenamefont {Dechant}\ and\ \citenamefont {Sasa}(2021)}]{dechant2021improving}%
  \BibitemOpen
  \bibfield  {author} {\bibinfo {author} {\bibfnamefont {A.}~\bibnamefont {Dechant}}\ and\ \bibinfo {author} {\bibfnamefont {S.-i.}\ \bibnamefont {Sasa}},\ }\bibfield  {title} {\bibinfo {title} {Improving thermodynamic bounds using correlations},\ }\href@noop {} {\bibfield  {journal} {\bibinfo  {journal} {Physical Review X}\ }\textbf {\bibinfo {volume} {11}},\ \bibinfo {pages} {041061} (\bibinfo {year} {2021})}\BibitemShut {NoStop}%
\bibitem [{\citenamefont {Bhatia}(2013)}]{bhatia2013matrix}%
  \BibitemOpen
  \bibfield  {author} {\bibinfo {author} {\bibfnamefont {R.}~\bibnamefont {Bhatia}},\ }\href@noop {} {\emph {\bibinfo {title} {Matrix analysis}}},\ Vol.\ \bibinfo {volume} {169}\ (\bibinfo  {publisher} {Springer Science \& Business Media},\ \bibinfo {year} {2013})\BibitemShut {NoStop}%
\end{thebibliography}%

\end{document}